\documentclass[12pt,preprint]{aastex}

\slugcomment{\today}

\shorttitle{Jellyfish candidates}
\shortauthors{Poggianti et al.}

\begin{document}


\title{Jellyfish galaxy candidates at low redshift}

\author{B.M. Poggianti$^1$, G. Fasano$^1$, A. Omizzolo$^{2,1}$,
  M. Gullieuszik$^1$, D. Bettoni$^1$, 
A. Moretti$^3$, A. Paccagnella$^{1,3}$, Y. L. Jaff\'e$^4$, B. Vulcani$^{5}$,
  J. Fritz$^6$, W. Couch$^7$, M. D'Onofrio$^3$}
\affil{$^1$INAF-Astronomical Observatory of Padova, Italy,
$^2$ Vatican Observatory, Vatican City State,
$^3$Physics and Astronomy Department, University of Padova, Italy,
$^4$Department of Astronomy, Universidad de Concepci\'on, Concepci\'on, Chile, $^5$Kavli Institute for the Physics and Mathematics of the
Universe (WPI), The University of Tokyo Institutes for Advanced Study
(UTIAS), the University of Tokyo, Kashiwa, 277-8582, Japan,
$^6$Centro de Radioastronom\'\i a y Astrof\'\i sica, CRyA, UNAM,
Michoac\'an, Mexico,$^7$Australian Astronomical Observatory, North Ryde, NSW 1670, Australia
}




\begin{abstract}
Galaxies that are being stripped of their gas can sometimes be
recognized from their optical appearance.
Extreme examples of stripped galaxies are the so-called ``jellyfish galaxies'', that exhibit
tentacles of debris material with a characteristic jellyfish morphology.
We have conducted the first systematic search for galaxies that are
being stripped of their gas
at low-z ($z=0.04-0.07$) in different environments, selecting galaxies
with varying degrees of morphological evidence for stripping.
We have visually inspected B and V-band
images and identified 344 candidates in 71 galaxy clusters of the OMEGAWINGS+WINGS sample and 75
candidates in groups and lower mass structures in the PM2GC sample.
We present the atlas
of stripping candidates and a first analysis of their
environment and their basic properties, such as morphologies, star
formation rates and galaxy stellar masses. 
Candidates are found in all clusters and at all
clustercentric radii,  and their number does not correlate with the cluster
velocity dispersion $\sigma$ or X-ray luminosity $L_X$. Interestingly,
convincing cases of candidates are also found in groups and lower
mass haloes ($10^{11}-10^{14} M_{\odot}$), although the physical
mechanism at work needs to be securely identified.
All the candidates are disky, have stellar masses ranging from 
$\rm log M/M_{\odot} < 9 $ to $> 11.5$
and the majority of them form stars
at a rate that is on average a factor of 2 higher (2.5$\sigma$)
compared to non-stripped galaxies of similar mass.
The few post-starburst and passive candidates have weak stripping evidence. 
We conclude that the stripping phenomenon is ubiquitous in clusters
and could be present even in groups and low mass haloes. Further studies will
reveal the physics of the gas stripping and clarify the mechanisms at work.
\end{abstract}





\section{Introduction}
In order to unveil the physical drivers of galaxy evolution,
it is crucial to study the processes of gas acquisition and loss.
Gas is the fuel for star formation (SF) and a sensitive tracer of environmental
effects.

Gas loss from galaxies can be caused by mechanisms internal to galaxies
themselves, such as galactic winds due to star formation or an AGN (e.g. Veilleux et
al. 2005, Ho et al. 2014, Fogarty et al. 2012).
In addition, several external mechanisms that can potentially impact on a galaxy gas
content have been proposed (Boselli \& Gavazzi 2006, De Lucia 2010).
Among those not directly affecting the galaxy stellar component,
there are ram pressure stripping from the disk due to the interaction between the galaxy 
interstellar medium (ISM) and the intergalactic medium (IGM,
Gunn \& Gott 1972), and
the removal of the hot gas halo surrounding the galaxy (the so-called
``strangulation'') either via ram pressure or via tidal stripping by the
halo potential
(Larson et al. 1980, Balogh et al. 2000).
While the first one partially or completely removes the ISM, the second one deprives the galaxy of its gas reservoir, and
leaves the existing ISM in the disk to be consumed by SF. 
Circumgalactic gas can also be shock-heated and, as a consequence, can stop
cooling in dark matter haloes above a critical mass (Dekel \& Birnboim 2006).
Other, less often cited, processes that can be as or even more efficient in
certain conditions are thermal evaporation (Cowie \&
Songaila 1977) and turbulent/viscous stripping (Nulsen 1982).
Among those processes that affect both gas and stars, instead, there
are strong tidal interactions and minor and major mergers
(expected to be more common in groups, Barnes \& Hernquist 1992, Mihos
\& Hernquist 1994), tidal effects of the cluster as a whole (Byrd \& Valtonen,
1990) and ``harassment'', i.e.
the cumulative effect of several weak and fast tidal encounters,
expected to be more efficient in galaxy clusters (Moore et al. 1996). 

Some of the most striking examples of gas stripping come from neutral
hydrogen studies. Neutral hydrogen gas has been observed to be disturbed and eventually truncated and 
exhausted in galaxies in dense environments, such as 
clusters (Davies \& Lewis 1973, Haynes et al. 1984, Giovanelli \&
Haynes 1985, Cayatte et al. 1990, Bravo-Alfaro et al. 2001,
Kenney et al. 2004, Chung et al. 2009, Jaff\'e et al. 2015) and groups (Williams \& Rood
1987, Rasmussen et al. 2006, Verdes-Montenegro et al. 2001, Sengupta \&
Balasubramanyam 2006, Rasmussen et al. 2008). 
These studies point to ram pressure stripping, or a combination of
ram pressure and tidal effects, as cause of the gas depletion.

Extreme examples of gas stripping are the so-called ``jellyfish
galaxies'' (e.g. Fumagalli et al. 2014, Ebeling et al. 2014). They exhibit
``tentacles'' of material that appear 
to be stripped from the main body of the galaxy, and whose morphology 
is suggestive of gas-only removal mechanisms, such as ram pressure
stripping.
Jellyfish galaxies (with different naming) have been known in nearby
clusters for many years. Usually, only a few galaxies
per cluster have been studied, in a handful of clusters (e.g. Virgo,
Coma, A1367, A3627, Shapley; Kenney \& Koopmann 1999, Sun et al. 2006,
Yoshida et al. 2008, Yagi et al. 2010, Smith
et al. 2010, Hester et al. 2010, Merluzzi et al. 2013, Kenney et al. 2014). A 
few examples have been identified in clusters at $z \sim 0.2-0.4$
(Owers et al. 2012, Ebeling et al. 2014, Rawle et al. 2014, Cortese et
al. 2007), and there is accumulating evidence for a
correlation between the efficiency of the stripping phenomenon
and the presence of shocks and strong gradients in the X-ray
IGM (Owers et al. 2012, Vijayaraghavan \& Ricker 2013). Known jellyfishes are star-forming or 
post-starburst galaxies; the existence of
ellipticals with X-ray tails might be a different side of the same coin
(e.g. Sun et al. 2005, Machacek et al. 2006).

$\rm H\alpha$ maps of jellyfish galaxies 
show tails of ionized gas up to 150 kpc long, where new
stars are born in knots and end up contributing to the intracluster
light.  A recent MUSE study of a jellyfish in a cluster
at z=0.016 has ruled out gravitational interactions as mechanism for the
gas removal and 
showed that ram pressure has removed the galaxy ISM
from the outer disk, while the primary $\rm H\alpha$ tail is still being fed
by gas from the galaxy inner regions (Fumagalli et al. 2014). 

The goal of this paper is to present the results of
a systematic search for galaxies whose optical morphology
suggests they might be experiencing stripping of their gaseous material.
By doing this, we aim to select all possible gas stripping candidates,
from the most extreme cases (with classical ``jellyfish'' morphology) to examples 
without obvious ``tentacles'' but with morphologies and/or surrounding
debris suggestive of stripping and/or ram pressure events.
This search has been conducted in galaxy clusters and in the general field at $z=0.04-0.07$, based on optical
images of the OMEGAWINGS+WINGS (Gullieuszik et
al. 2015, Fasano et al. 2006) and PM2GC (Calvi et al. 2011) samples
described in sec.~2. The process and criteria for candidate selection is presented
in sec.~3, and the methods to derive star formation rates and
morphologies in sec.~3.1.
This paper presents the atlas of
images and the catalogs (\S4), the environments (\S5), the morphologies
and stellar population properties (star formation rates, stellar masses, colors
and spectral types) of 419 stripping candidates (\S6).

In this paper we use $\Omega_{m}$=0.3, $\Omega_{\Lambda}$=0.7, $\rm H_0=70 \, km \,s^{-1}Mpc^{-1}$
and a Kroupa (2001) IMF.

\section{Datasets}

\subsection{WINGS and OMEGAWINGS}

WINGS is a large survey targeting 76 clusters of
galaxies selected on the basis of their X-ray luminosity (Ebeling et al. 1996, 1998,
2000), 
covering a wide range in cluster masses ($\sigma=$
500-1200+ $\rm km \, s^{-1}$, $\rm log \, L_X =
43.3-45 \, \rm erg \, s^{-1}$, Fasano et al. 2006). 
The original WINGS dataset consisted of B
and V deep photometry of a  $34^{\prime} \times 34^{\prime}$
field-of-view with the WFC@INT and
the WFC@2.2mMPG/ESO (Varela et al. 2009), spectroscopic follow-ups with 2dF@AAT and
WYFFOS@WHT (Cava et al. 2009), plus J and K imaging with WFC@UKIRT
(Valentinuzzi et al. 2009) and some U-band imaging (Omizzolo et al. 2014). This
database is presented in Moretti et al. (2014)
and has been employed for a number of studies (see https://sites.google.com/site/wingsomegawings/).

OMEGAWINGS is a recent extention of this project, that quadruples the
area covered (1 square degree) and allows to reach up to $\sim$2.5
cluster virial radii. OMEGAWINGS is based on two OmegaCAM@VST 
GTO programs for 46 WINGS clusters: a B and V campaign completed
in P93, and an ongoing u-band programme. The B
and V data, the data reduction and the photometric catalogs are
presented in Gullieuszik et al. (2015). Spectra are
obtained with AAOmega@AAT on the OmegaCAM field. 
So far, we have secured high quality
spectra for $\sim 30$ OMEGAWINGS clusters, reaching very high 
spectroscopic completeness levels for
galaxies brighter than V=20 from the cluster cores to their periphery
(Moretti et al. in prep.). Galaxies are considered cluster members if
they are within 3$\sigma$ from the cluster redshift. The mean redshift
uncertainty, computed from the differences between WINGS and
OMEGAWINGS redshift values of repeated objects, is $\Delta z = 0.0002$.

For this paper we consider the 41 OMEGAWINGS clusters with an OmegaCAM
B and/or V-band seeing $\leq 1.2$arcsec, listed in Table~1. 
Due to the segmentation of the B and V OmegaCAM filters, the OmegaCAM images
have a central vignetting cross (Gullieuszik et al. 2015): only in
the vignetted area we used the old WINGS images (Fasano et al. 2006).
Finally, to complete the search within the WINGS sample, we used
the old WINGS images for other 31 clusters not observed
with OmegaCAM (Table~1). In the following, we keep these clusters separate from the rest
because the WINGS imaging covers only the cluster cores (the central $\sim
0.3$ sq. deg.). The masses of OMEGAWINGS and
WINGS clusters have been estimated from the $\sigma$ applying the
virial theorem according to eqn.~4 in Poggianti et al. (2006).

\begin{table}
\scriptsize
\caption{OMEGAWINGS (O) and WINGS (W) clusters. }
\begin{center}
\begin{tabular}{cllccc}
Sample &Cluster & redshift & $\sigma$ & $Log(L_X)$ & $N_{candidates}$ \\
&&& km/s& 0.1-2.4keV& \\
O &  A1069 &   0.0651 &677  $\pm$    52    &   43.98    &	  2	 \\		  
O & A119  &   0.0444  &862  $\pm$    52    &   44.51    &	  7	 \\		 
O & A147  &   0.0447  &666  $\pm$    13    &   43.73    &	  8\\			 
O &  A151  &   0.0532  &760  $\pm$    55    &   44.00    &	  5\\			 
O &  A160  &   0.0438  &561  $\pm$    53    &   43.58    &	  5\\			 
O &  A1631a*&   0.0465 &750  $\pm$    28    &   43.86    &	  4 \\			 
O &  A168  &   0.0448  &503  $\pm$    43    &   44.04    &	  6\\			 
O &  A193  &   0.0484 &777  $\pm$    72    &   44.19    &	  1  \\    			 
O &  A1983 &   0.0447  &527  $\pm$    38    &   43.67    &	  3    \\ 			 
O &   A1991 &   0.0584  &599  $\pm$    57    &   44.13    &	  6     \\			 
O &   A2107 &   0.0410  &592  $\pm$    62    &   44.04    &	  4	\\		 
O &   A2382 &   0.0639 &699  $\pm$    30    &   43.96    &	  2    	\\		 
O &   A2399 &   0.0577 &722  $\pm$    35    &   44.00    &	  3    	\\		 
O &   A2415 &   0.0575  &696  $\pm$    51    &   44.23    &	  7  	\\		 
O &   A2457 &   0.0587 &685  $\pm$    36    &   44.16    &	  3  	\\		 
O &   A2589 &   0.0419  &816  $\pm$    88    &   44.27    &	  5       \\ 		 
O &   A2593 &   0.0417  &701  $\pm$    60    &   44.06    &	  6  	\\		 
O &   A2657 &   0.0400 &381  $\pm$    83    &   44.20    &	  2     \\			 
O &   A2665 &   0.0562  &---                 &   44.28    &	  1   	\\		 
O &   A2734 &   0.0624  &555  $\pm$    42    &   44.41    &	  2       \\		 
O &   A3128 &   0.0603  &854  $\pm$    28    &   44.33    &	  4  	\\		 
O &   A3158 &   0.0594 &997  $\pm$    38    &   44.73    &	  4      \\			 
O &   A3266 &   0.0595 &1309 $\pm$    39    &   44.79    &	  1       \\		 
O &   A3395 &   0.0507 &1206 $\pm$    55    &   44.45    &	  4        \\		 
O &   A3528*&   0.0535  &899 $\pm$     64    &   44.12    &	  4 \\
O &   A3530*&   0.0548 &700 $\pm$     40    &   43.94    &	  5   \\			 
O &   A3532*&   0.0555  &621 $\pm$     53    &   44.45    &	  4     \\   		 
O &   A3556*&   0.0480  &640 $\pm$     35    &   43.97    &	  1 \\
O &   A3558*&   0.0485 &1049$\pm$     52    &   44.80    &   14 \\
O &   A3560*&   0.0489  &710 $\pm$     41    &   44.12    &	  8\\
O &   A3667 &   0.0558 &1011 $\pm$    42    &   44.94    &	  2   \\   			 
O &   A3716 &   0.0457 &842  $\pm$    27    &   44.00    &	  2    	\\		 
O &   A3809 &   0.0626 &558  $\pm$    38    &   44.35    &	  2   	\\		 
O &   A3880 &   0.0585 &531  $\pm$    35    &   44.27    &	  7  	\\		 
O &   A4059 &   0.0480  &715  $\pm$    59    &   44.49    &	  4       \\  		 
O &   A754  &   0.0545 &1039 $\pm$    63    &   44.90    &	  3     \\			 
O &  A85   &   0.0521  &1052 $\pm$    68    &   44.92    &	  3       \\  		 
O &  A957x &   0.0451  &710  $\pm$    53    &   43.89    &	  3      \\			 
O & IIZW108&   0.0487 &617  $\pm$    42    &   44.34    &	  2     \\			 
O &  MKW3s &   0.0444  &539  $\pm$    37    &   44.43    &	  1   	\\		 
O &  Z8852 &   0.0408  &765  $\pm$    63    &   43.97    &    2      \\                  
W  & A133 & 0.0603 & 810$\pm$78 & 44.55 & 5 \\  
W  & A311 & 0.0657 & --- & 43.91 & 2 \\  
W  & A376 & 0.0476 & 852$\pm$49 & 44.14 & 1 \\  
W  & A500 & 0.0678 & 658$\pm$48 & 44.15 & 4 \\  
W  & A602 & 0.0621 & 720$\pm$73 & 44.05 & 1 \\  
W  & A671 & 0.0507 & 906$\pm$58 & 43.95 & 1 \\  
W  & A1291 & 0.0509 & 429$\pm$49 & 43.64 & 1 \\  
W  & A1644 & 0.0467 & 1080$\pm$54 & 44.55 & 15 \\  
W  & A1668 & 0.0634 & 649$\pm$57 & 44.20 & 8 \\  
W  & A1736 & 0.0461 & 853$\pm$60 & 44.37 & 14 \\  
W  & A1795 & 0.0633 & 725$\pm$53 & 45.05 & 9 \\  
W  & A1831 & 0.0634 & 543$\pm$58 & 44.28 & 1 \\  
W  & A2124 & 0.0666 & 801$\pm$64 & 44.13 & 6 \\  
W  & A2149 & 0.0675 & 353$\pm$53 & 43.92 &  3\\  
W  & A2169 & 0.0578 & 509$\pm$40 & 43.65 & 4 \\  
W  & A2256 & 0.0581 & 1273$\pm$64 & 44.85 & 4 \\  
W  & A2572a & 0.0390 & 631$\pm$10 & 44.01 & 1 \\  
W  & A2622 & 0.0610 & 696$\pm$55 & 44.03 &  1 \\  
W  & A2626 & 0.0548 & 625$\pm$62 & 44.29 & 3\\  
W  & A2717 & 0.0498 & 553$\pm$52 & 44.00 & 1 \\  
W  & A3164 & 0.0611 & --- & 44.17 & 2 \\  
W  & A3376 & 0.0461 & 779$\pm$49 & 44.39 & 2 \\  
W  & A3490 & 0.0688 & 694$\pm$52 & 44.24 & 4\\  
W  & A3497 & 0.0680 & 726$\pm$47 & 44.16 & 4 \\  
W  & RX0058 & 0.0484 & 637$\pm$97 & 43.64 & 3 \\  
W  & RX1022 & 0.0548 & 577$\pm$49 & 43.54 & 1 \\  
W  & RX1740 & 0.0441 & 582$\pm$65 & 43.70 & 3 \\  
W  & Z1261 & 0.0644 & --- & 43.91 & 4 \\  
W  & Z2844 & 0.0503 & 536$\pm$53 & 43.76 & 1 \\  
W  & Z8338 & 0.0494 & 712$\pm$60 & 43.90 & 2 \\  
\end{tabular}
\tablecomments{Name, redshift, $\sigma$, $L_X$ and number of 
  stripping candidates (spectroscopic members or possible members,
  i.e. without redshift) of OMEGAWINGS and WINGS clusters. Those with an asterisk belong 
  to the Shapley supercluster.}
\end{center}
\end{table}


\subsection{PM2GC}
The reference comparison field sample for WINGS and OMEGAWINGS is
the Padova Millennium Galaxy and Group Catalogue (PM2GC, Calvi et al. 2011),
built from  the Millennium Galaxy Catalogue (MGC, Liske et al. 2003), 
a deep 38 $deg^2$ INT B-imaging survey with a highly complete
spectroscopic follow-up (96\%  at B=20, Driver et al. 2005). 
The image quality (for depth, pixel size, median seeing)
and the spectroscopic completeness of the PM2GC are superior to Sloan, and
these qualities result in more robust morphological
classifications and better sampling of dense regions. 
This, and the fact that the observational data is very similar to WINGS
and was analyzed in the same way with the same tools,
make the PM2GC  the ideal field counterpart for WINGS. 
The characterization of the environment of PM2GC 
galaxies was conducted by means of a Friends-of-Friends 
algorithm by Calvi et al. (2011), who identified 176 groups of
galaxies with at least three members\footnote{A galaxy is considered
  member of a group if its spectroscopic redshift lies within 3$\sigma$ 
from the median group redshift and if it is located within a projected
distance of 1.5$R_{200}$ from the group geometrical center. $R_{200}$ is defined as the radius 
  delimiting a sphere with interior mean density 
200 times the critical density of the universe at that 
redshift. $R_{200}$ is 
commonly used as an approximation of the group/cluster virial radius and is 
computed from the group or cluster velocity dispersion as in Poggianti et al. (2006).}
as well as binary systems and ``single'' galaxies, respectively defined as
galaxies with just one or no neighbor with a projected mutual
distance of $ \leq 0.5$ $h^{-1}$ Mpc and a redshift within $1500 \rm \, km \, s^{-1}$. 
The masses of the dark matter haloes hosting
PM2GC galaxies were estimated from the correlation
between the dark matter halo mass and the total stellar mass of
member galaxies (Paccagnella et al. in
prep).


\section{Data analysis}

Two or three of us (first independently, then together) visually inspected the OMEGAWINGS and PM2GC (GF and BP) and
WINGS-only (AO, GF and BP) B-band
images searching for galaxies with optical evidence for gas stripping.\footnote{For OMEGAWINGS,
the V-band image was used if the median B-band seeing for that cluster
was worse than 1.2arcsec.} 
We searched for any type of evidence suggestive of gas stripping,
selecting galaxies that have a) debris trails, tails or surrounding 
debris located on one side of the galaxy and/or b) asymmetric/disturbed 
morphologies suggestive of unilateral external forces,
and/or c) a distribution of star-forming regions and knots
suggestive of triggered star formation on one side of the galaxy.
These criteria are similar to those used in previous studies of
jellyfish galaxies (e.g. Ebeling et al. 2014).

For OMEGAWINGS+WINGS we inspected the whole image of each
cluster, and inspected all galaxies in the image, without knowing
their magnitude and whether they had a measured redshift. This
selection yielded candidates with a B-band Sextractor AUTO magnitude
(corrected for Galactic extinction) $\leq 20$ (Gullieuszik et al. 2015).
For the PM2GC, instead, we looked at all galaxies with a spectroscopic
redshift in the same range of WINGS clusters ($z=0.04-0.07$), thus
starting from the spectroscopic MGC catalog that has a limit B=20.
Thus, the classification process was done blindly with respect to
environment: the classifier knew neither whether the galaxy was a
cluster/group member, nor the distance to the cluster or group
center. He/she only knew whether the image was from OMEGAWINGS, WINGS
or PM2GC.

The images were first inspected independently by each classifier, who
assigned a class according to the scheme described below.
The different classifiers agreed to within one class in 83\% of the cases.
Then, each galaxy was inspected together by the common classifiers to ensure
homogeneity, the final classification was agreed upon and a consensus
was found on the classification of those
galaxies whose individual class differed by more than one class.

We tentatively assigned our candidates to five classes 
according to the visual evidence for stripping signatures in the optical bands
 (JClass), from extreme cases 
(JClass 5) to progressively weaker cases, down to the weakest (JClass 1).
As a result, our JClass=5 and 4 classes comprise the most secure candidates,
and contain the most striking classical jellyfish galaxies. JClass 3
candidates are probable cases of stripping and/or ram pressure events,
while JClasses 2 and 1 are tentative candidates, for which definitive
conclusions cannot be reached on the basis of the existing imaging.

To this concern, however, it is interesting to note that integral-field spectroscopy of one of our weakest-class 
candidates (JClass 1) clearly shows one-sided extraplanar ionized gas 
stripped by ram pressure (Merluzzi et al. 2013). This galaxy, shown
at the bottom of Fig.~1, was selected by our visual inspection
as a JClass=1 and was later recognized as the galaxy studied by
Merluzzi et al. (see figures 5,6,7 in their paper for the outstanding
stripped $\rm H\alpha$-emitting gas on one side of it).
In general, spatially resolved gas-sensitive studies of
galaxies in the process of being stripped have shown that 
the optical signatures are just the tip of the iceberg ({e.g. Fumagalli
et al. 2014, Merluzzi et al. 2013, Kenney et al. 2015)}:
the ongoing stripping is much more evident from the ionized gas
than in the optical. This leads us to suspect that stripping takes place
even in the optically weakest cases, and it is the reason why we deemed
it useful to include in our catalog galaxies over the whole range of degree of
evidence for stripping.

Instead, we deliberately tried to remove from the catalog galaxies with morphologies
clearly disturbed due to mergers or tidal
interactions, still retaining and flagging the most doubtful
cases where either gas stripping or tidal forces, or both, might be at
work. In fact, the eventual presence of tidal forces does not exclude the
possibility that also gas stripping mechanisms, such as ram
pressure, are at work, as it is sometimes observed  (e.g. NGC 4654 in
Virgo, Vollmer 2003). 
Thus, the reader should be aware that the catalog comprises
galaxies for which the optical morphology alone is not sufficient to
identify beyond any doubt the physical origin of the stripping, that can
be pinpointed only by gas-sensitive follow-up studies. Only subsequent
studies, in fact, will be able to discriminate between different
processes that can give origin to similar morphological features, such
as harassment (especially in clusters, Moore et al. 1996) and minor
mergers (also in groups and low density environments, Bournaud et
al. 2004, Cox et al. 2008, Hopkins et al. 2009, Lotz et al. 2010a, 2010b).

It is important to keep in mind that the ``JClass'' 
depends not only on the intrinsically stronger or weaker evidence for 
stripping signatures, but also on the galaxy orientation with respect 
to the line of sight, the galaxy size (number of pixels) and the 
signal-to-noise of the images,
thus it is only crudely indicative of the effective degree of 
surrounding debris.
 
\subsection{Galaxy properties}

The galaxy current star formation rate (SFR), stellar mass and
absolute magnitudes were 
derived applying our spectrophotometric tool SINOPSIS to the available optical
spectroscopy, as described in
Fritz et al. (2011) for WINGS, Moretti et al. (in prep.) for OMEGAWINGS and Poggianti et al. (2013) for PM2GC.
The model performs a non-parametric full spectral fitting of the
continuum shape and of the main emission and absorption lines,
deriving a star formation history. The ongoing SFR is constrained from the
fluxes of the emission lines and the blue part of the spectrum, and
dust extinction is taken into account (see Fritz et al. 2007, 2011 for details).
Being obtained from multifiber spectroscopy, the SFR estimate
refers to the central region of galaxies that is covered by the fiber
(that has a diameter of 2.1arcsec, covering the central 1.7-2.8
kpc at the WINGS redshifts), and is then extrapolated to a total SFR value
assuming the same mass-to-light ratio within and outside of the fiber.
The mean correction factor is 6, with a standard deviation of 4.8.

Galaxies were assigned a spectral type on the basis of the strength of
their main emission and absorption lines, as done in Fritz et
al. (2014). In the following we distinguish between ``star-forming
galaxies'' (all those with emission lines) and ``post-starburst'' (k+a) and
``passive'' (k) galaxies that lack emission lines and have a strong
and a weak $\rm H\delta$ line in absorption, respectively. The
post-starburst signature testifies the presence of recent star
formation activity that ended sometime during the last $\sim 1$ Gyr,
while k galaxies are those that have been devoid of any star formation
for a longer time (Poggianti et al. 1999, Fritz et al. 2014).

Morphological classifications are available for WINGS (Fasano et
al. 2012) and PM2GC galaxies (Calvi et al. 2012).\footnote{The
  morphological analysis of the OMEGACAM images is ongoing, therefore,
  for now,  for the OMEGAWINGS clusters morphologies are only available on the
  smaller area covered by WINGS.}
They were obtained with MORPHOT, an automated tool designed to reproduce as
closely as possible the visual classifications (Fasano et al. 2012). MORPHOT adds to the
classical CAS (concentration/asymmetry/clumpiness) parameters a set of
additional indicators derived from digital imaging of galaxies and has
been proved to give an uncertainty similar to that of
eyeball estimates. It uses a combination of a
Neural Network and a Maximum Likelihood technique assigning a
morphological class T with a scale resembling the Revised
Hubble Type classification, for which 
T=-5 = elliptical, -2 = S0, 1 = Sa, 3 = Sb, 5 = Sc, 7 = Sd, 9 = Sm.


\section{Atlas}

Table~2 presents the number of candidates in the three samples,
globally and for each JClass separately, as well as the number of 
candidates with an available spectroscopic redshifts. 
In total, our sample comprises 419 candidates, of which
10 of JClass=5, 24 of JClass=4, 73 of JClass=3, 143 of JClass=2 and
169 of JClass=1.

Figures~1 to 5 present illustrative examples of 
OMEGAWINGS, WINGS and PM2GC candidates.
We present a single filter image with two different cuts, as well as a color-composite image.

For OMEGAWINGS, in Fig.~1 we show all JClass=5 galaxies, six JClass=4
candidates and one example of each of JClass=3,2 and 1. 
For WINGS, we show examples of each JClass in Fig.~2.
Finally, for PM2GC we show two examples for each
JClass from 4 to 1 in Fig.~3
(no JClass=5 candidate is present in the PM2GC).

The complete atlas with all images in pdf format 
is available in the online version of the paper, where we give both
the rgb image and two individual filter images if more than one filter
is available (for OMEGAWINGS and WINGS). 
In addition, for the JClass=1 OMEGAWINGS candidates, which are the hardest to  
visualize, we provide two pdf images, with different cuts.
Since the appearance
of the pdf figures strongly depends on the screen or printer used and
this can make it hard to visualize the features of interest, we also provide
OMEGAWINGS cutouts images of each candidate in fits format to allow the reader to
display each image with the most appropriate cuts for her/his
screen/printer. For WINGS, all the images are public through the VO, as described in
detail in Moretti et al. (2014).
For the PM2GC, all the images are made public by
the Millennium Galaxy Catalogue team (http://www.eso.org/$\sim$jliske/mgc/).

The full version of Tables~3 and 4 containing the catalogs of all 
candidates is available online. These tables list positions on the sky, 
JClass and relative comments, redshift when available and the redshift
source. For OMEGAWINGS
and WINGS, the eventual membership to the cluster and the type of
image and filter used are also given. The comments include some notes
on the several peculiar morphologies we have encountered, for example 
describe shapes similar to a ``croissant'', or to a bicycle's ``handlebar'', or
if a galaxy looks comet-like/tadpole.

Whenever a candidate could be suspected to be tidally interacting, due
to presence of a nearby galaxy and/or to a possibly winding morphology of the
tails, or to have possibly experienced a minor mergers, this
possibility has been recorded in the comments
as ``tidal''. Similarly,
we have recorded whether the morphological signature might resemble
the one expected from the harassment process. About 20\%
of candidates in OMEGAWINGS and 40\% in WINGS and PM2GC have been flagged as possibly
tidal, interacting, merging or harassed.  
Instead, as explained in \S3, we tried not to include in the sample galaxies
with clear evidence for a tidal interaction or a merger.

This is by far the largest existing sample of stripping candidates, with
344 galaxies in OMEGAWINGS+WINGS and 75 in the PM2GC
sample, and a spectroscopic redshift for $\sim 70$\% of them. 
They are homogeneously selected and cover a wide range of
environments, that will be discussed in the following section.

\begin{table}
\caption{Number of stripping candidates.}
\begin{center}
\begin{tabular}{lllccllclccl}
Sample  &   $N_{stripping}$ & $N_5$ & $N_4$ & $N_3$ & $N_2$ & $N_1$ & $N_{z}$  & $N_{mem}$   & $N_{other-mem}$  & $N_{non-mem}$  \\ 
&&&&&&& \\
OMEGAWINGS & 211 & 8 & 19 & 48 & 66 & 70 & 156  & 107 & 7 & 42 \\ 
WINGS  &  133 & 2 & 2 &19 & 49 & 61 & 77 & 55 & 0 & 22 \\
PM2GC  & 75 & 0 & 3 & 6 & 28 & 38 & 75 & -- & -- & -- \\ 
Total & 419 & 10 & 24 & 73 & 143 & 169 & 308 & nd & nd & nd \\
\end{tabular}
\tablecomments{Columns: (1) Sample, (2) number of candidates,
  (3,4,5,6,7) number of JClass=5,4,3,2,1 candidates, (8) number of
  candidates with a known spectroscopic redshift, (9) number of
  spectroscopic cluster members, (10) number of spectroscopic members
  of other known structures in the fore/background of the main cluster, (11)
  number of galaxies with redshift that belong neither to the main
  cluster nor to other fore/background known structures.}
\end{center}
\end{table}


\begin{figure*}
\vspace{-3.2cm}
\centerline{\includegraphics[scale=0.8]{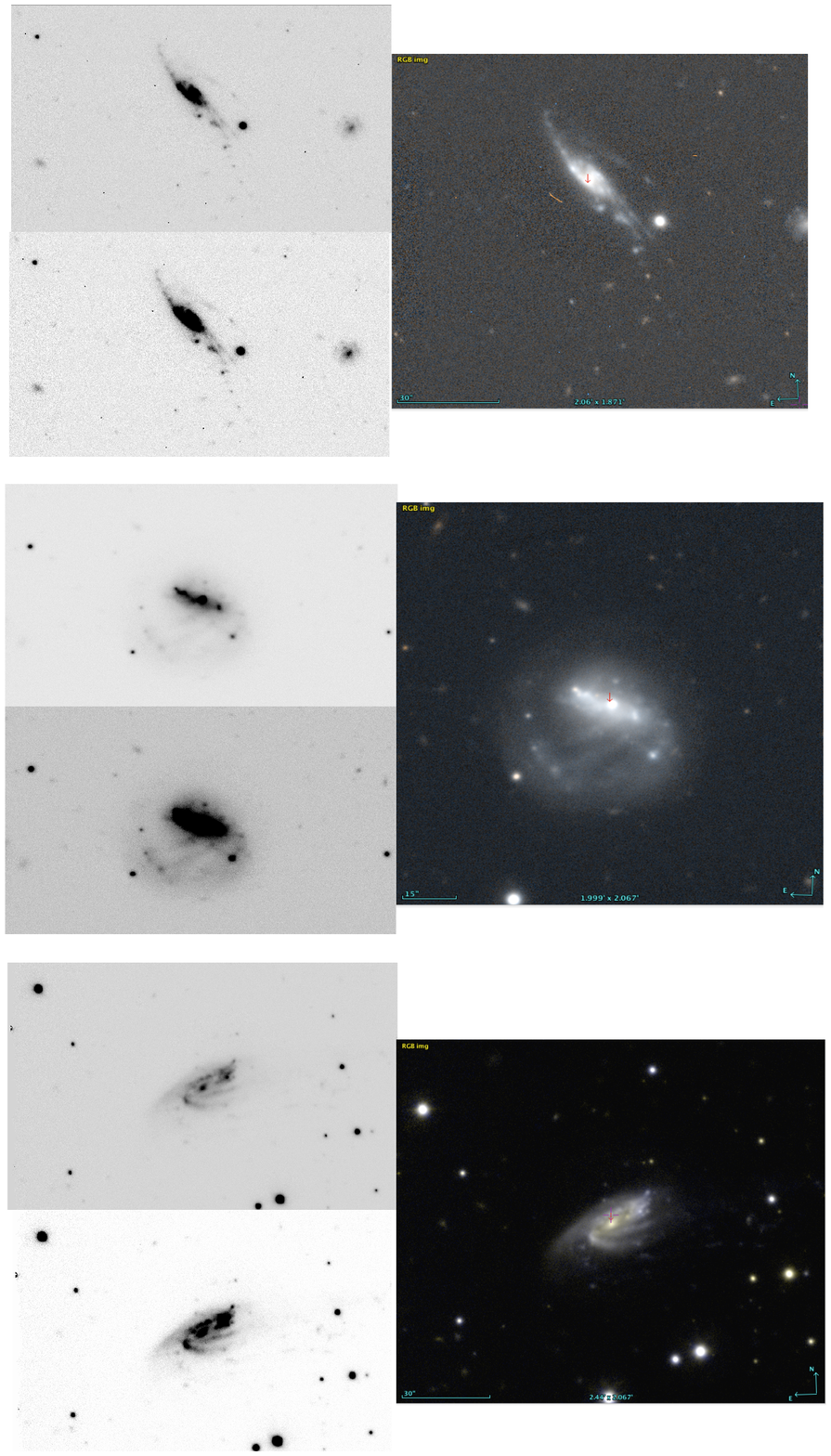}}
\vspace{-1.0cm}
\caption{JClass=5 OMEGAWINGS stripping candidates. {\bf Left} Single
  filter images with two different level cuts. 
 {\bf Right} Color-composite image.
}
\end{figure*}

\begin{figure*}
\vspace{-2.8cm}
\centerline{\includegraphics[scale=0.9]{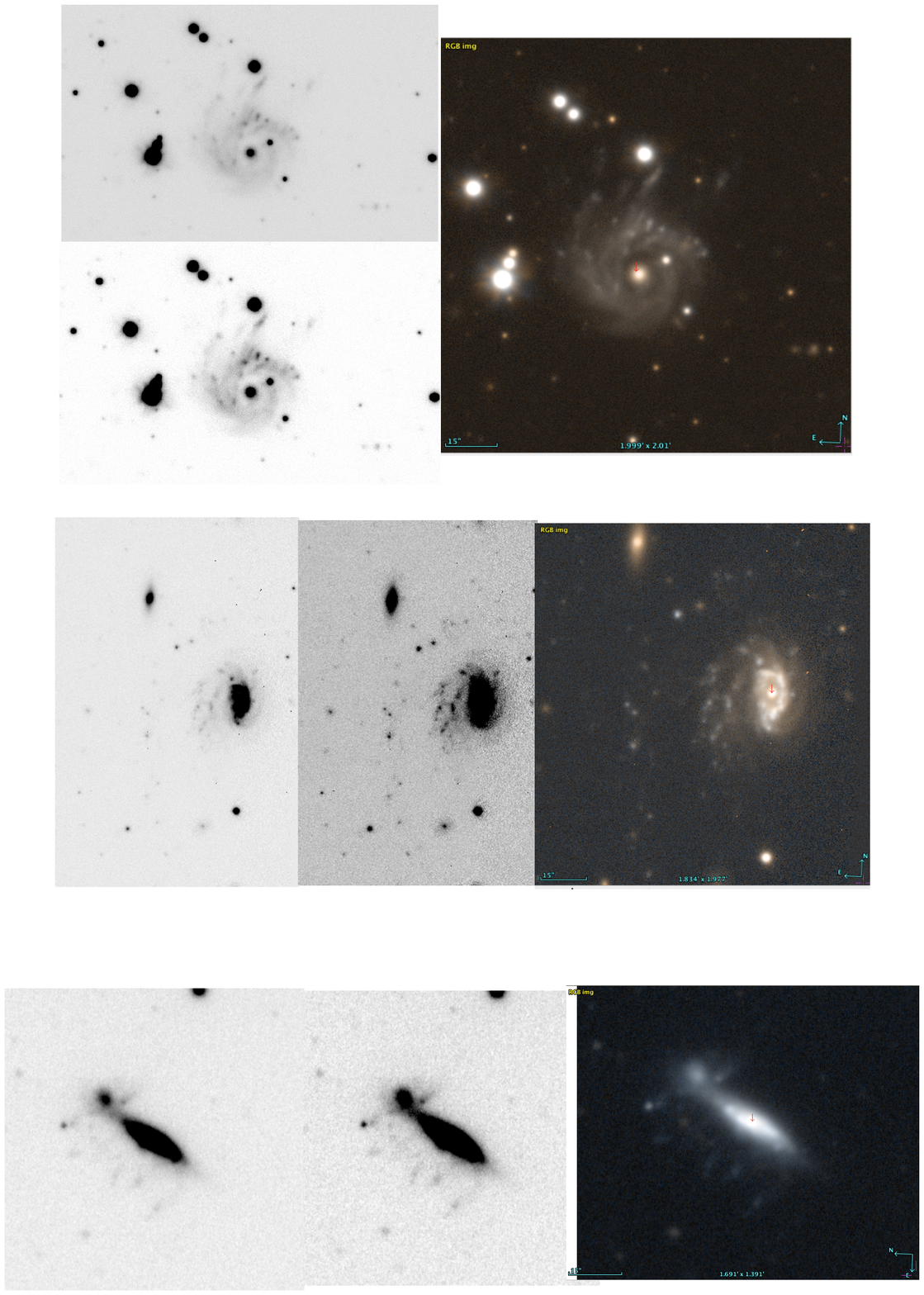}}
\vspace{-2.5cm}
\setcounter{figure}{0}
\caption{continues. JClass=5 OMEGAWINGS candidates. 
}
\end{figure*}

\begin{figure*}
\vspace{-3cm}
\centerline{\includegraphics[scale=0.8]{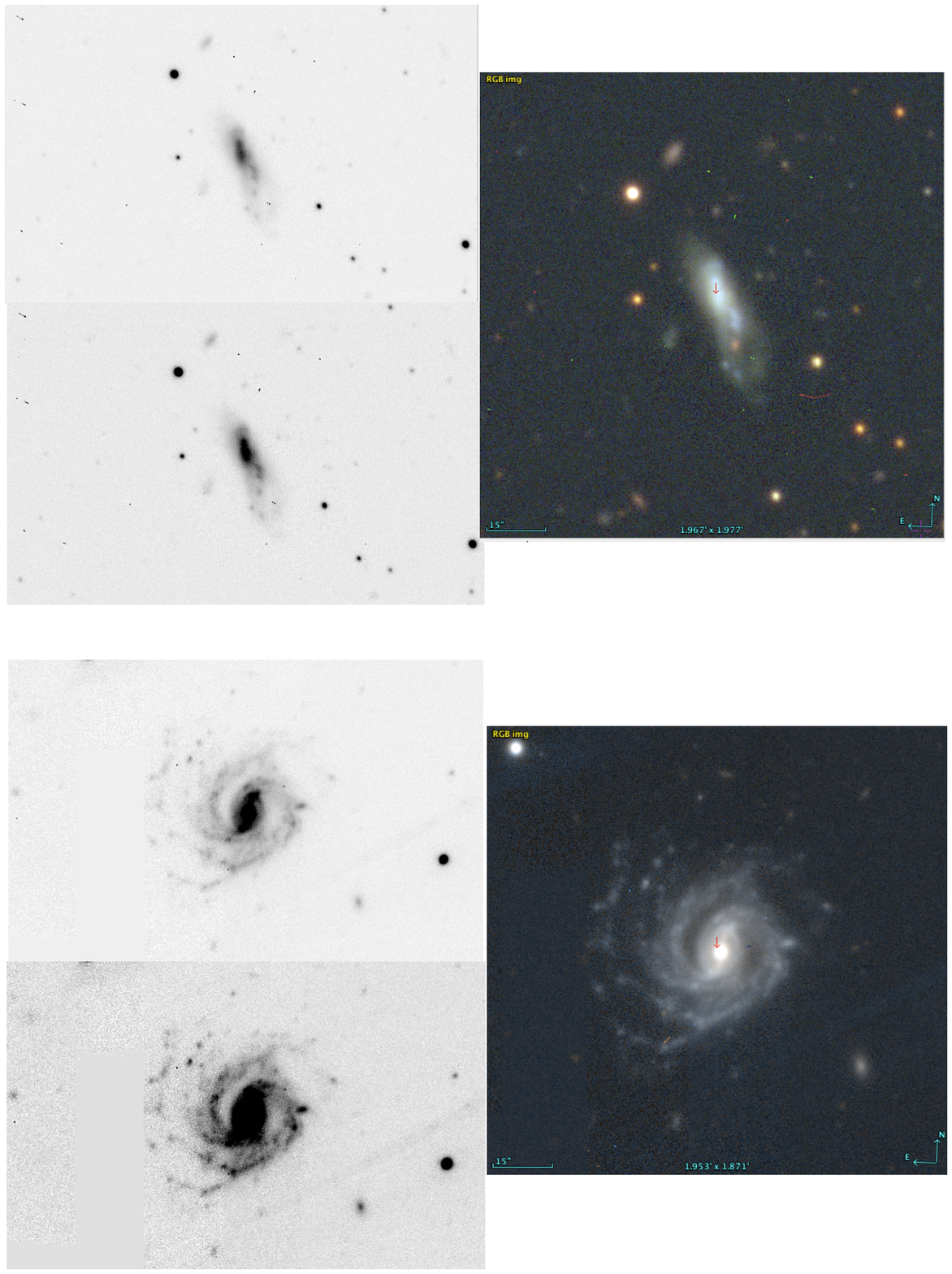}}
\vspace{-3cm}
\setcounter{figure}{0}
\caption{continues. JClass=5 OMEGAWINGS candidates. 
}
\end{figure*}

\begin{figure*}
\vspace{-3cm}
\centerline{\includegraphics[scale=0.9]{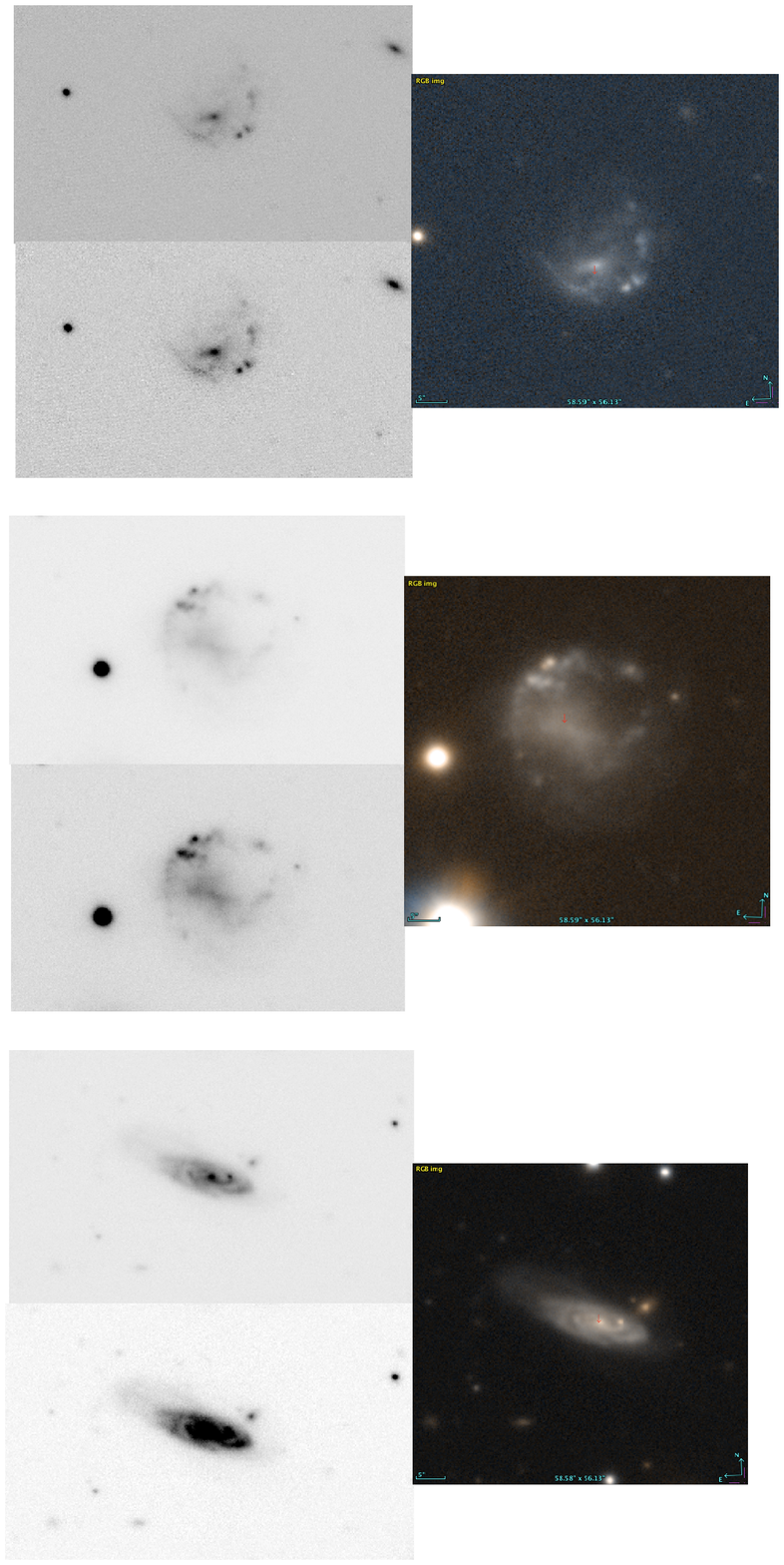}}
\vspace{-2cm}
\setcounter{figure}{0}
\caption{continues. Examples of  JClass=4  OMEGAWINGS candidates. 
}
\end{figure*}

\begin{figure*}
\vspace{-3cm}
\centerline{\includegraphics[scale=0.83]{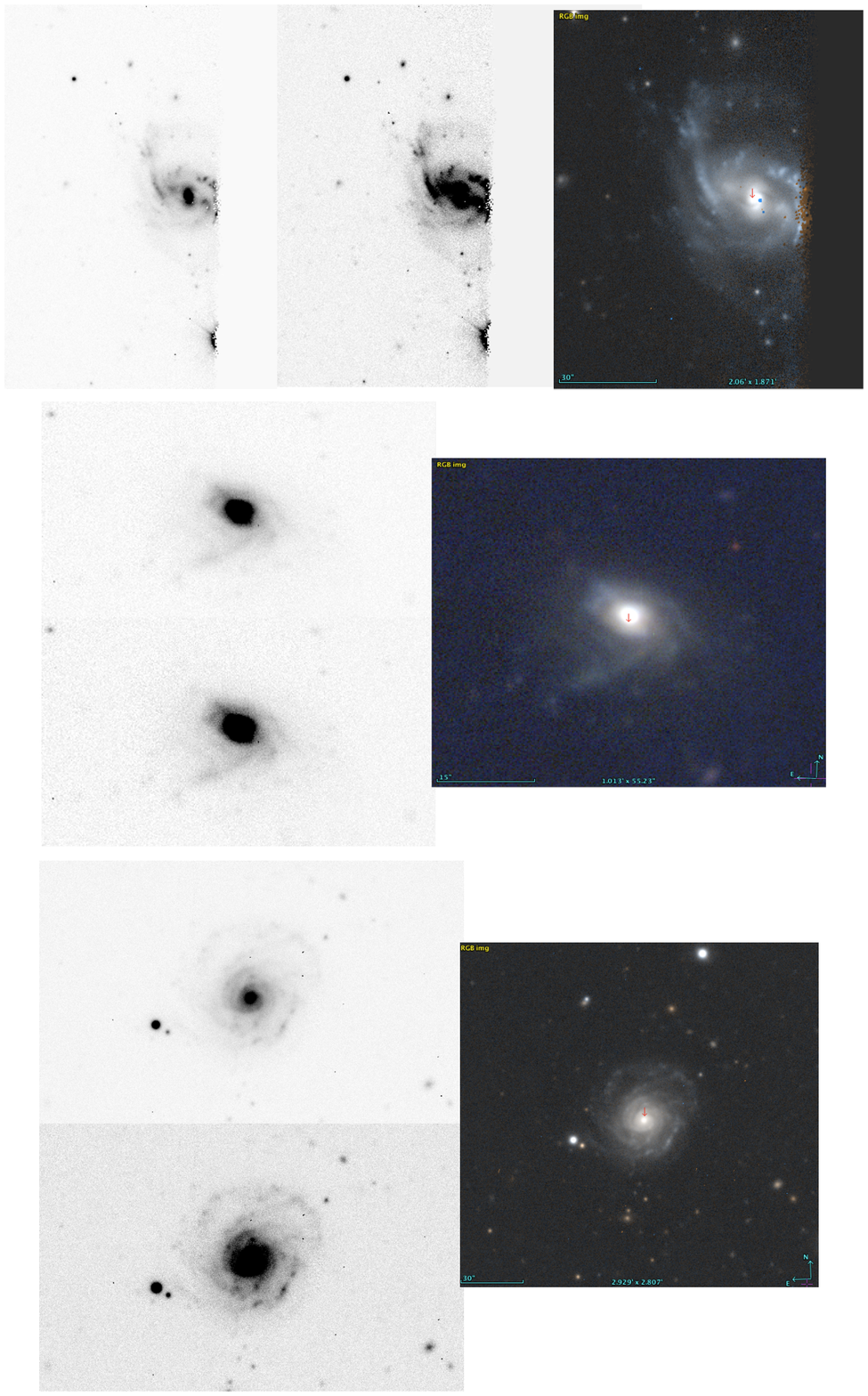}}
\vspace{-1cm}
\setcounter{figure}{0}
\caption{continues. Examples of  JClass=4  OMEGAWINGS candidates. 
}
\end{figure*}

\begin{figure*}
\vspace{-3cm}
\centerline{\includegraphics[scale=0.8]{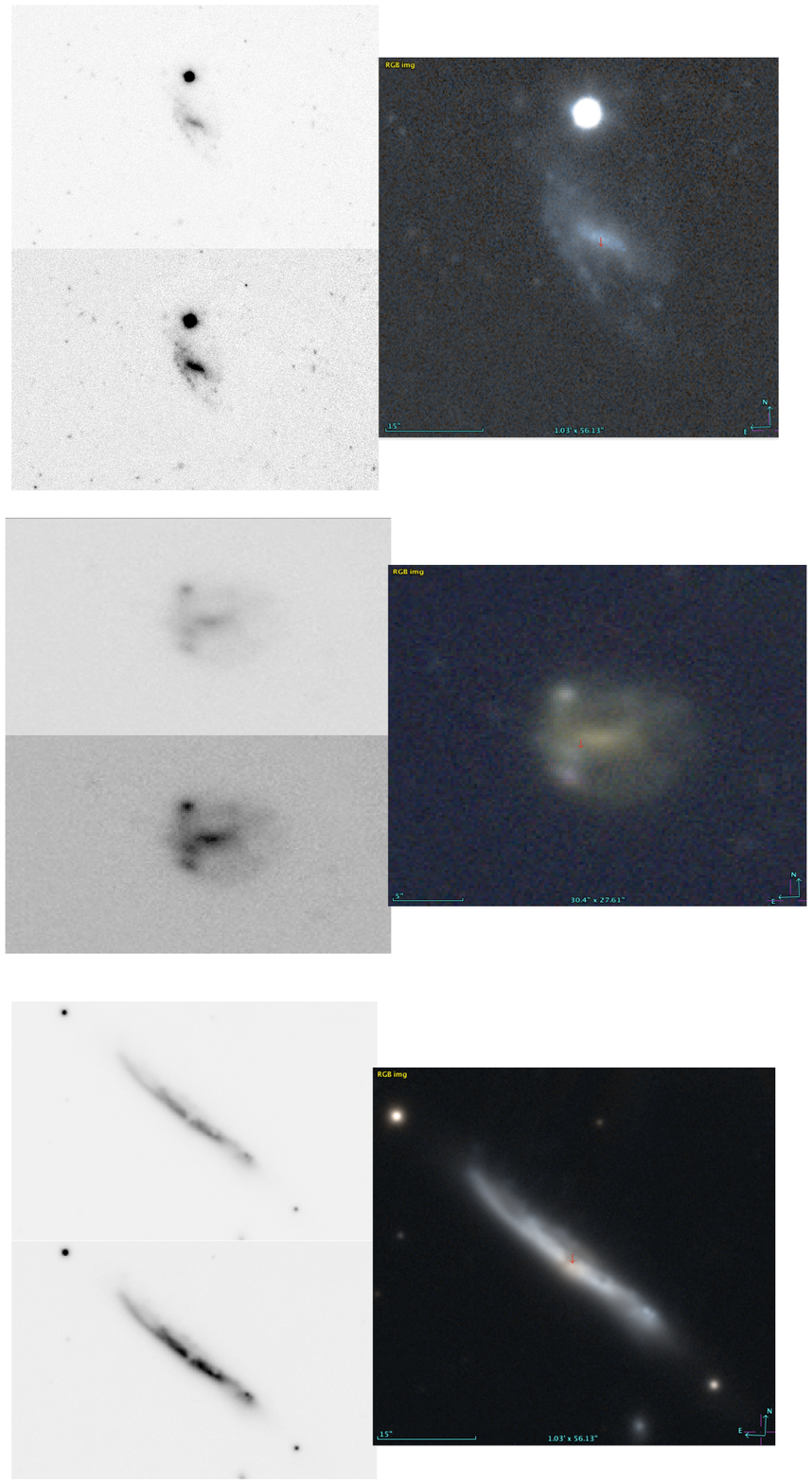}}
\vspace{-2cm}
\setcounter{figure}{0}
\caption{continues. Examples of JClass=3 (top),  JClass=2 (middle) and JClass=1
  (bottom) OMEGAWINGS candidates. The JClass=1 galaxy is the one
  studied in Merluzzi et al. (2013).
}
\end{figure*}

\begin{figure*}
\centerline{\includegraphics[scale=0.8]{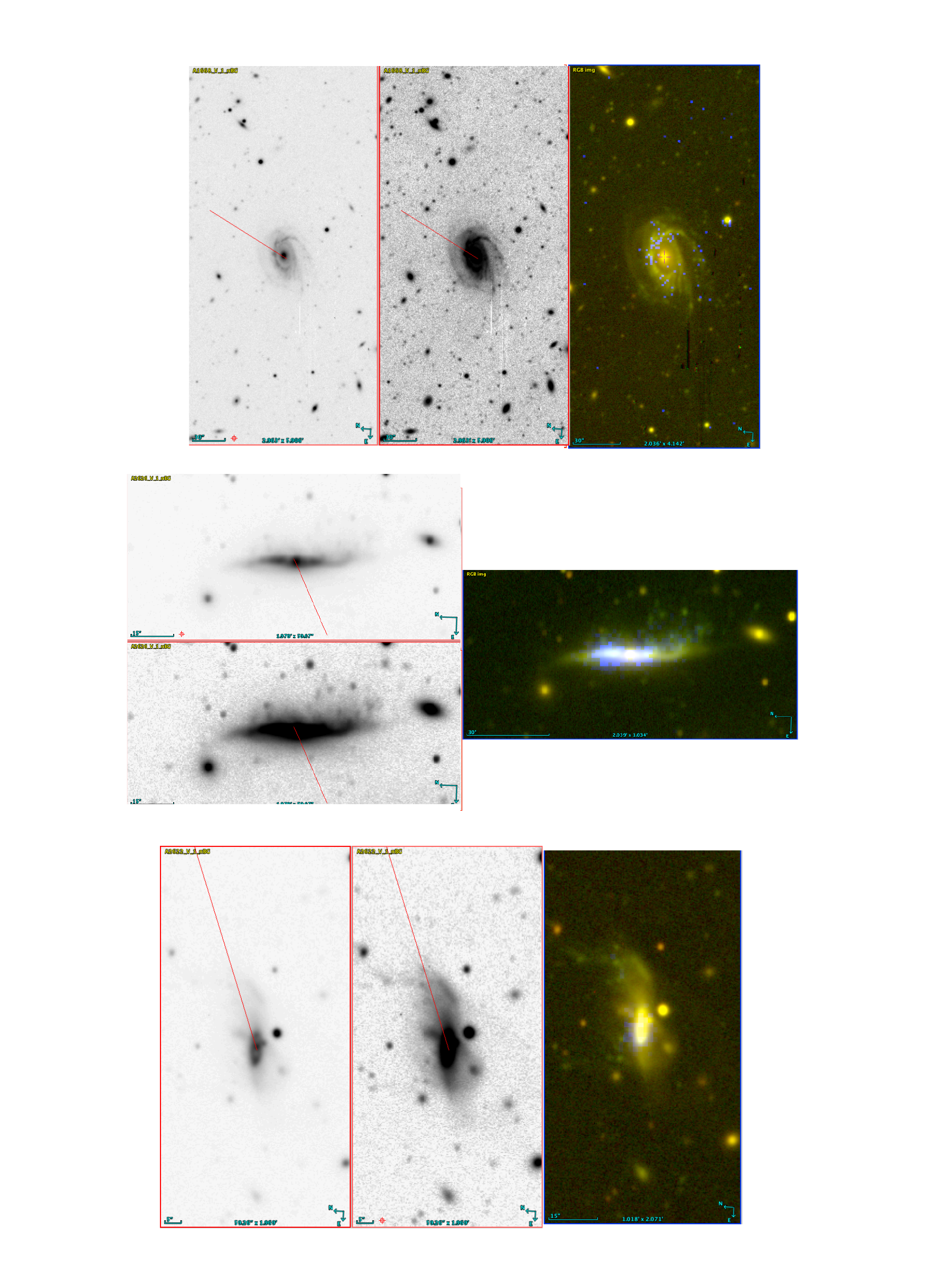}}
\setcounter{figure}{1}
\caption{Examples of JClass=5 (top and middle) and JClass=4 (bottom) WINGS candidates. 
}
\end{figure*}

\begin{figure*}
\vspace{-2cm}
\centerline{\includegraphics[scale=0.8]{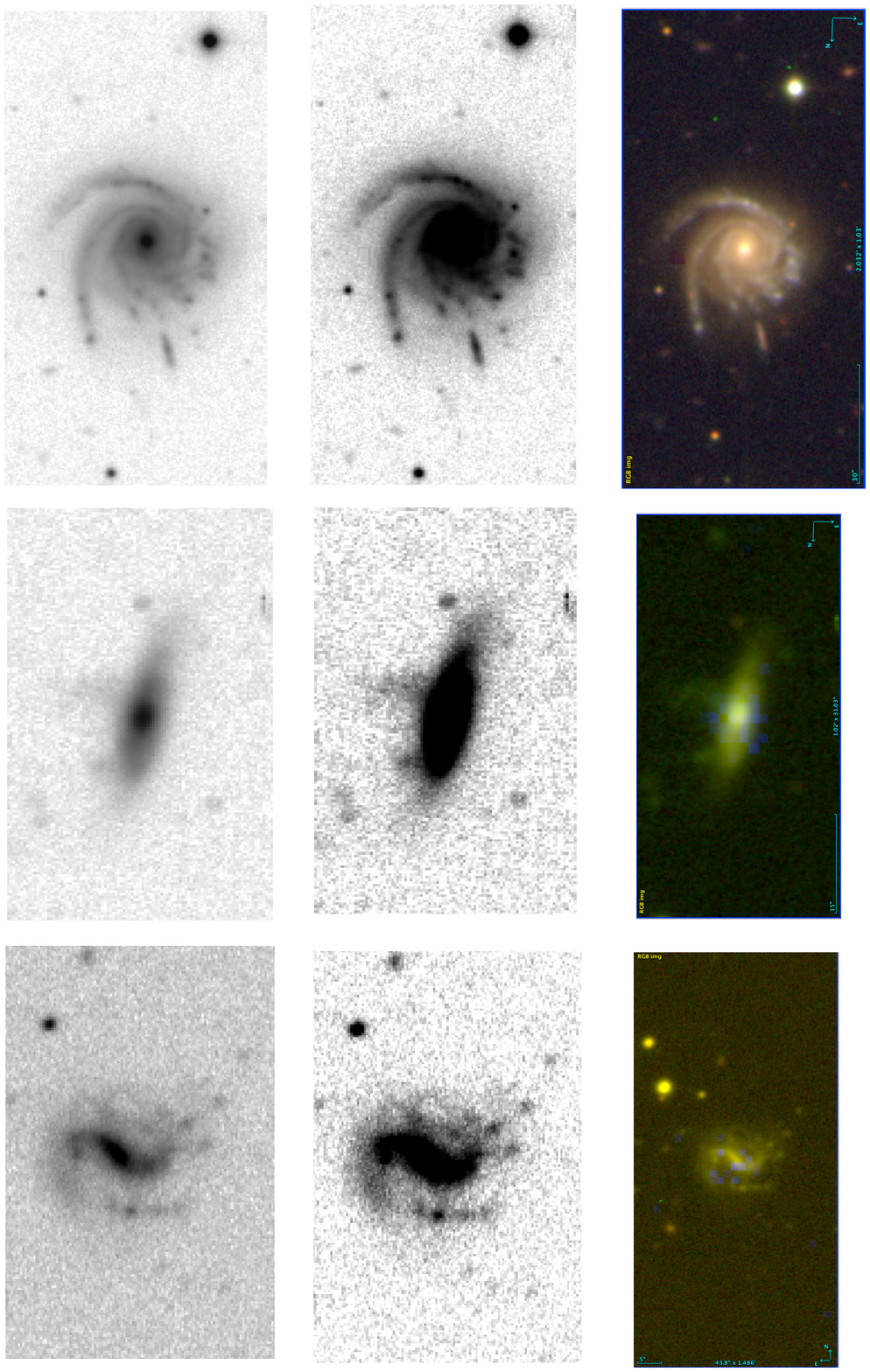}}
\vspace{-2cm}
\setcounter{figure}{1}
\caption{continues. Examples of JClass=3 (top), JClass=2 (middle) and JClass=1
  (bottom) WINGS candidates. Left and center: B-band image with two
  different stretches. Right: color-composite image.
}
\end{figure*}

\begin{table*}
\scriptsize
\caption{OMEGAWINGS and WINGS stripping candidates.}
\begin{center}
\begin{tabular}{llccllclccl}
ID  & Cluster & Image & band & RA & DEC & JClass & comments &
                                                                 Memb. &
                                                                         z
  & Source z \\
&&&&&&&&&& \\
  JO1         &      A1069    &      O      &      V      &       160.4327677   &       -8.4198195    &       1       &       disturbed     &       2       &       0.05765   &       1    \\  
  JO2         &      A1069    &      O      &      V      &       160.1087159   &       -8.2662606    &       2       &       -             &       -1      &       -----     &       ---    \\
  JO3         &      A1069    &      O      &      V      &       160.1466322   &       -8.4630044    &       2       &       -             &       2       &       0.15894   &       1     \\
  JO4         &      A1069    &      O      &      V      &       159.9728672   &       -8.9068175    &       2       &       -             &       2       &       0.05541   &       1      \\
  JO5         &      A1069    &      O      &      V      &       160.334885    &       -8.8961043    &       3       &       -             &       1       &       0.06484   &        1     \\
  JO6         &      A119     &      W      &      B      &       14.2420529    &       -1.2992219    &       1       &       tidal         &       0       &       0.07326   &       1      \\
  JO7         &      A119     &      O      &      B      & 13.806715
                                  &       -1.0760668    &       2 &       -
                                                            &       1
                                                                       &
                                                                         0.04762
  &        3   \\
  JO8       &      A119     &      O      &      B      &       14.4873539    &       -1.3355165    &       2       &       -             &       -1      &       -----     &       ---    \\
.... & .... & .... & .... & .... & .... & .... & .... & .... & .... & .... \\
  JW1 & A133 & W & B & 15.5496061 & -21.6593958 & 3 &  tidal & -1 & -- & --\\
  JW2 & A133 & W & B & 15.7618501 & -21.6613987 & 3 & warping & -1 & -- & --\\
  JW3 & A133 & W & B & 15.8219828 & -21.7460609 & 2 &  -- & 1 & 0.0529 & 2\\
  JW4 & A133 & W & B & 15.4267328 & -21.9509146 & 1 &  -- & -1 & -- & --\\
  JW5 & A133 & W & B & 15.5625195 & -22.0113691 & 1 &  -- & 1 & 0.0515 & 2\\
  JW6 & A311 & W & B & 32.2268083 & 19.7558379 & 1 &  tidal & -1 & -- & --\\
  JW7 & A311 & W & B & 32.0011882 & 19.698108 & 2 &  tidal & -1 & -- & --\\
\end{tabular}
\tablecomments{Columns: (1) ID: JO for OMEGAWINGS clusters and JW for
  WINGS clusters (2) Cluster (3) Image inspected:
  O=OMEGAWINGS (OmegaCAM) or W=WINGS (INT or 2.2m) (4) Band used (5)
  RA (J2000) (6) DEC (J2000) (7) JClass: from 5 (strongest)
  to 1 (weakest) (8) comments (9) Cluster membership: 1=member, 0=non
  member, 2= member of structure along the line of sight, -1=
  redshift unknown (10) Redshift (11) 
Source of redshift: 1=WINGS (Cava et al. 2009) or OMEGAWINGS (Moretti et al. in prep.),  2=NED, 3=SDSS, 4=average NED 
  and SDSS. 
The table is published in its entirety in the 
 electronic edition of the journal.  A portion is 
 shown here for guidance regarding its form and content.}
\end{center}
\end{table*}

\begin{figure*}
\vspace{-0.5cm}
\centerline{\includegraphics[scale=0.34]{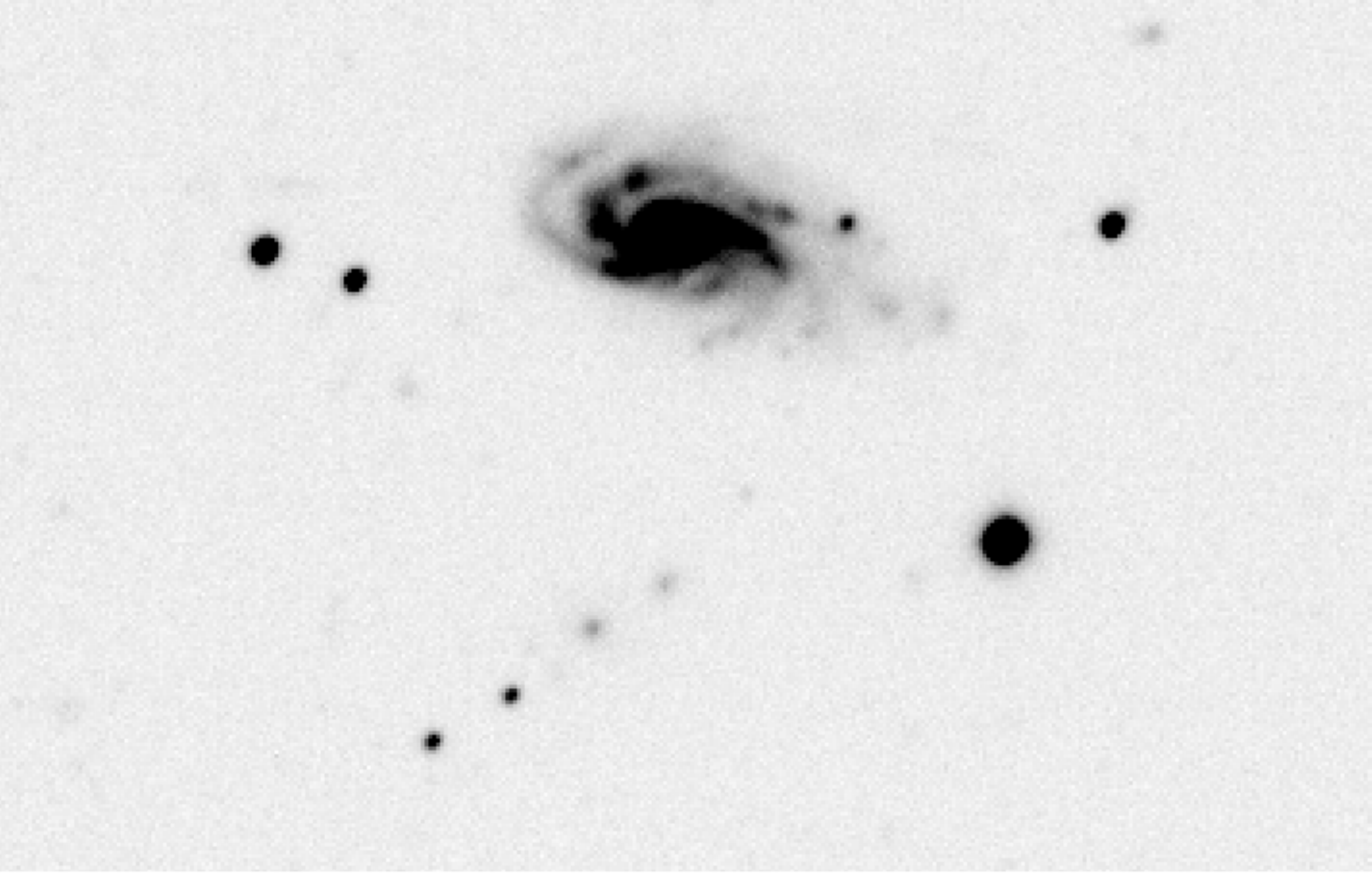}\hfill\hspace{-0.4cm}\includegraphics[scale=0.34]{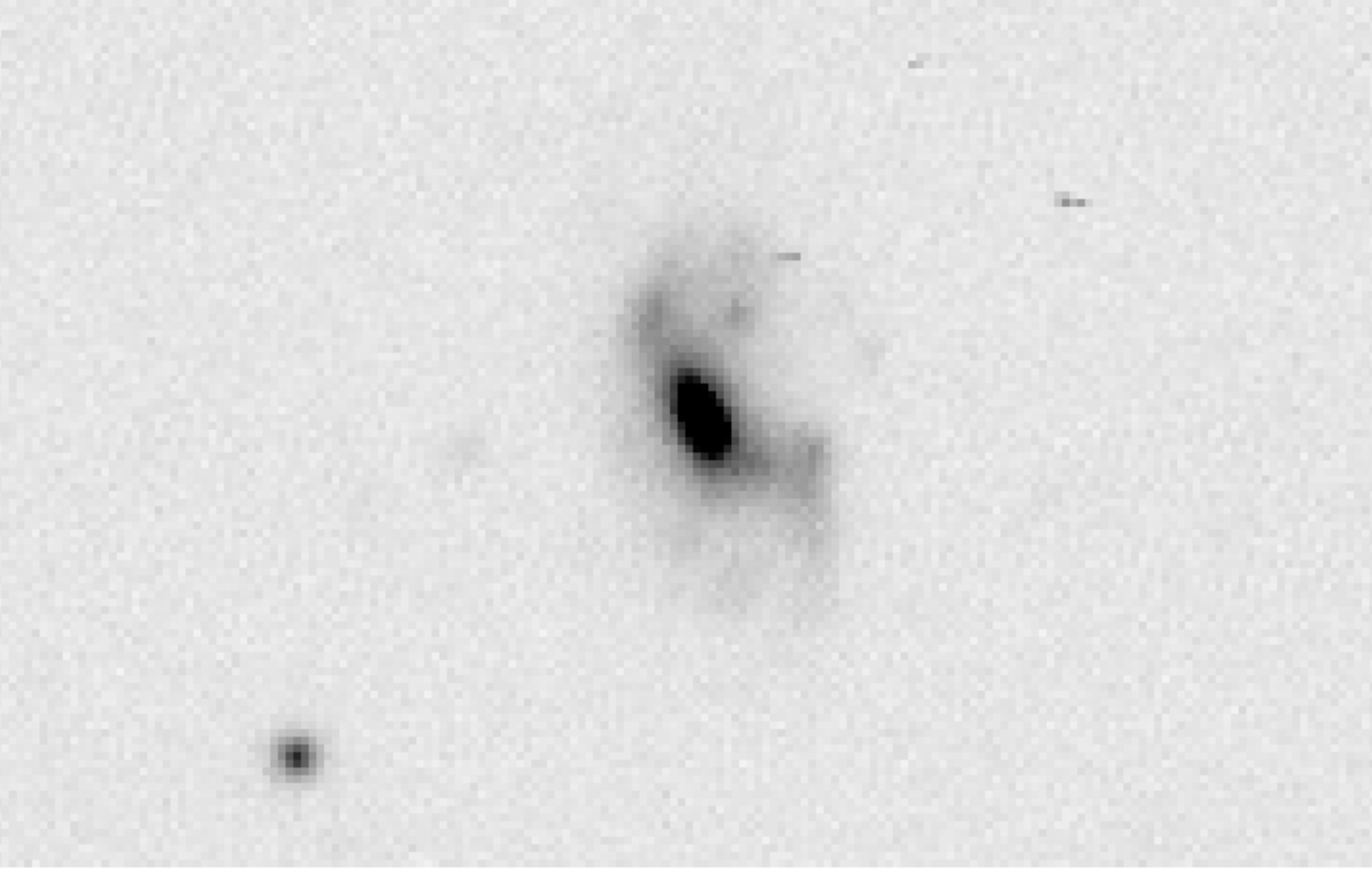}}
\centerline{\includegraphics[scale=0.34]{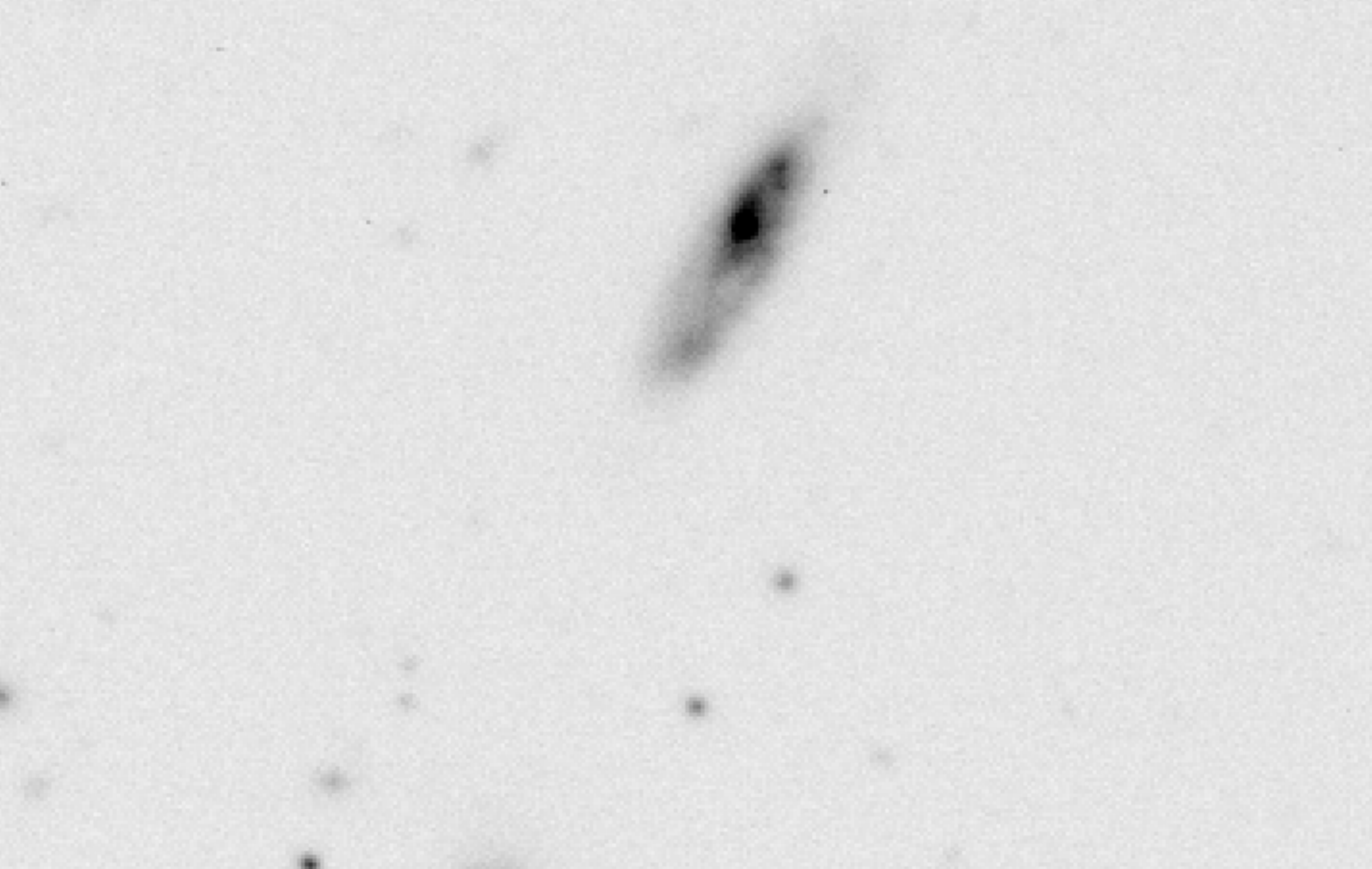}\hfill\hspace{-0.4cm}\includegraphics[scale=0.34]{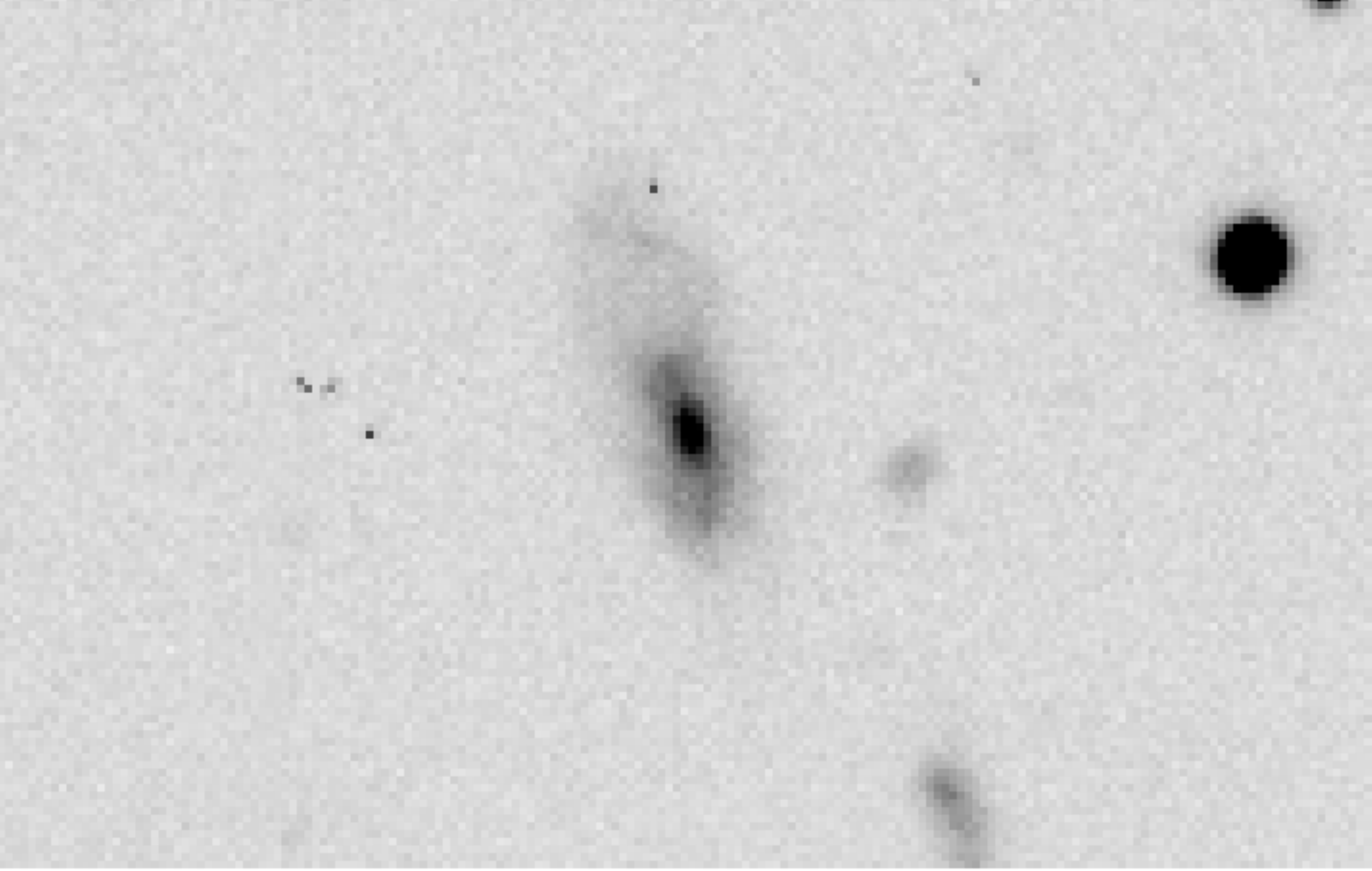}}
\centerline{\includegraphics[scale=0.34]{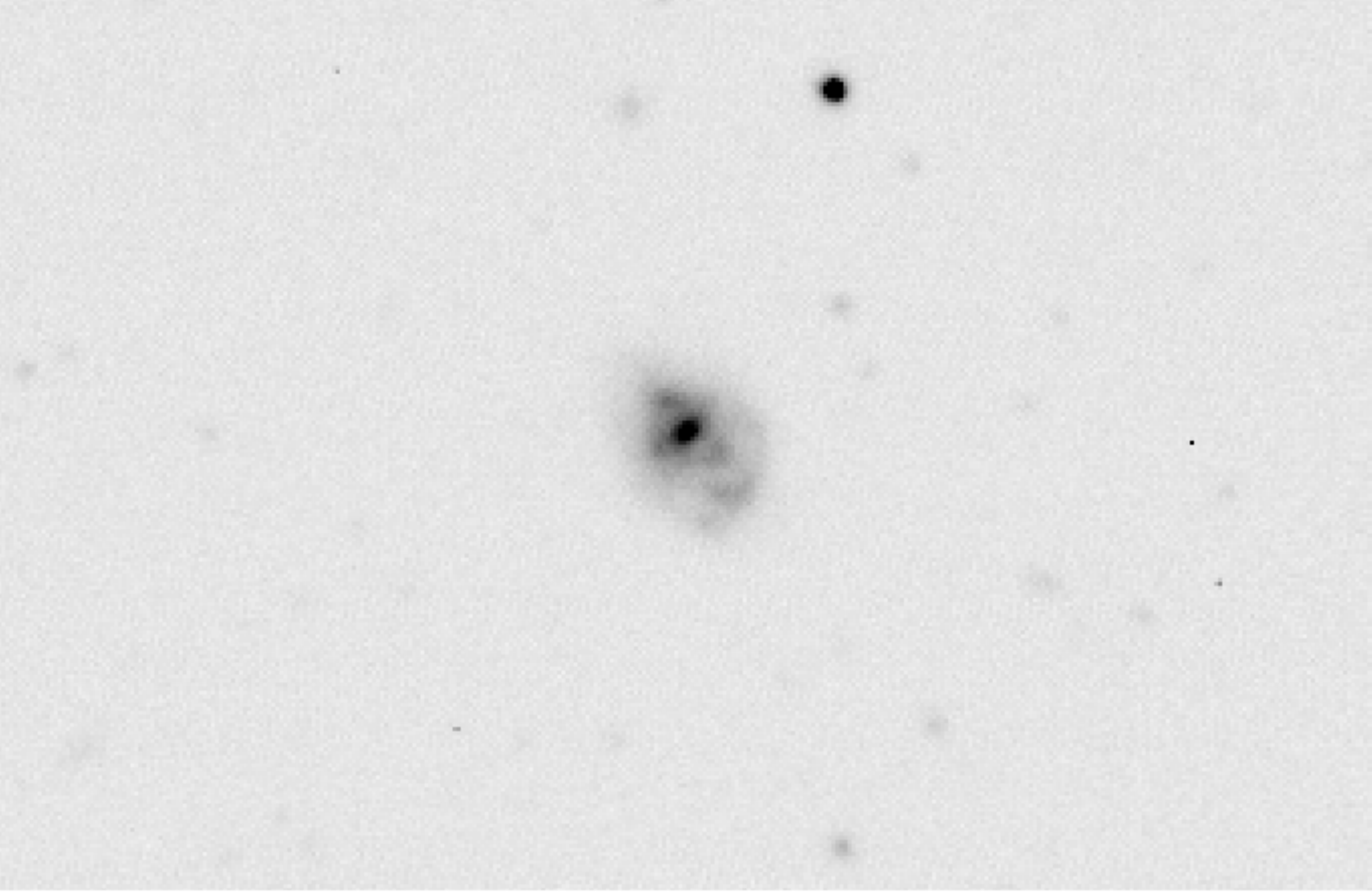}\hfill\hspace{-0.4cm}\includegraphics[scale=0.34]{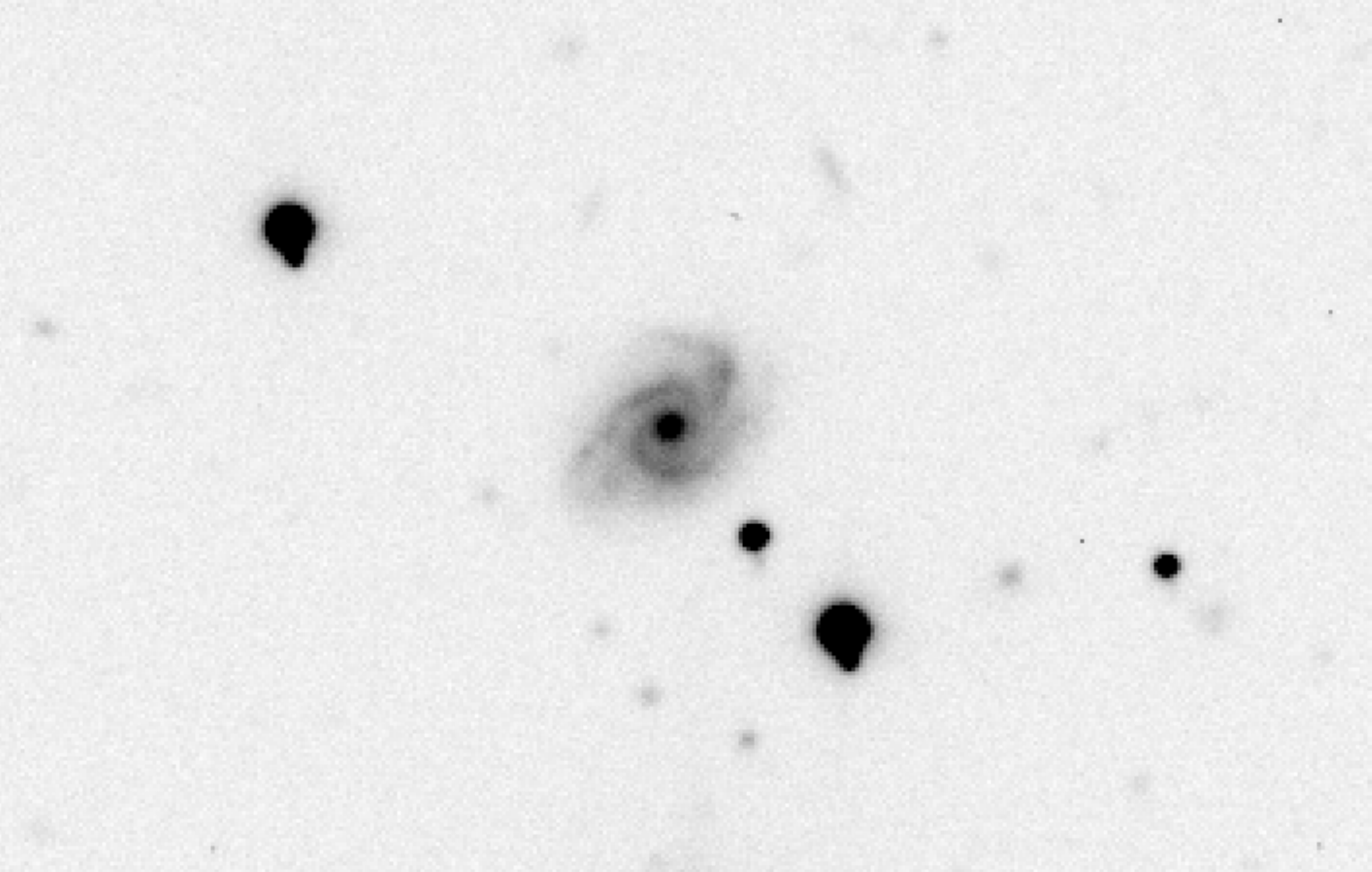}}
\centerline{\includegraphics[scale=0.3]{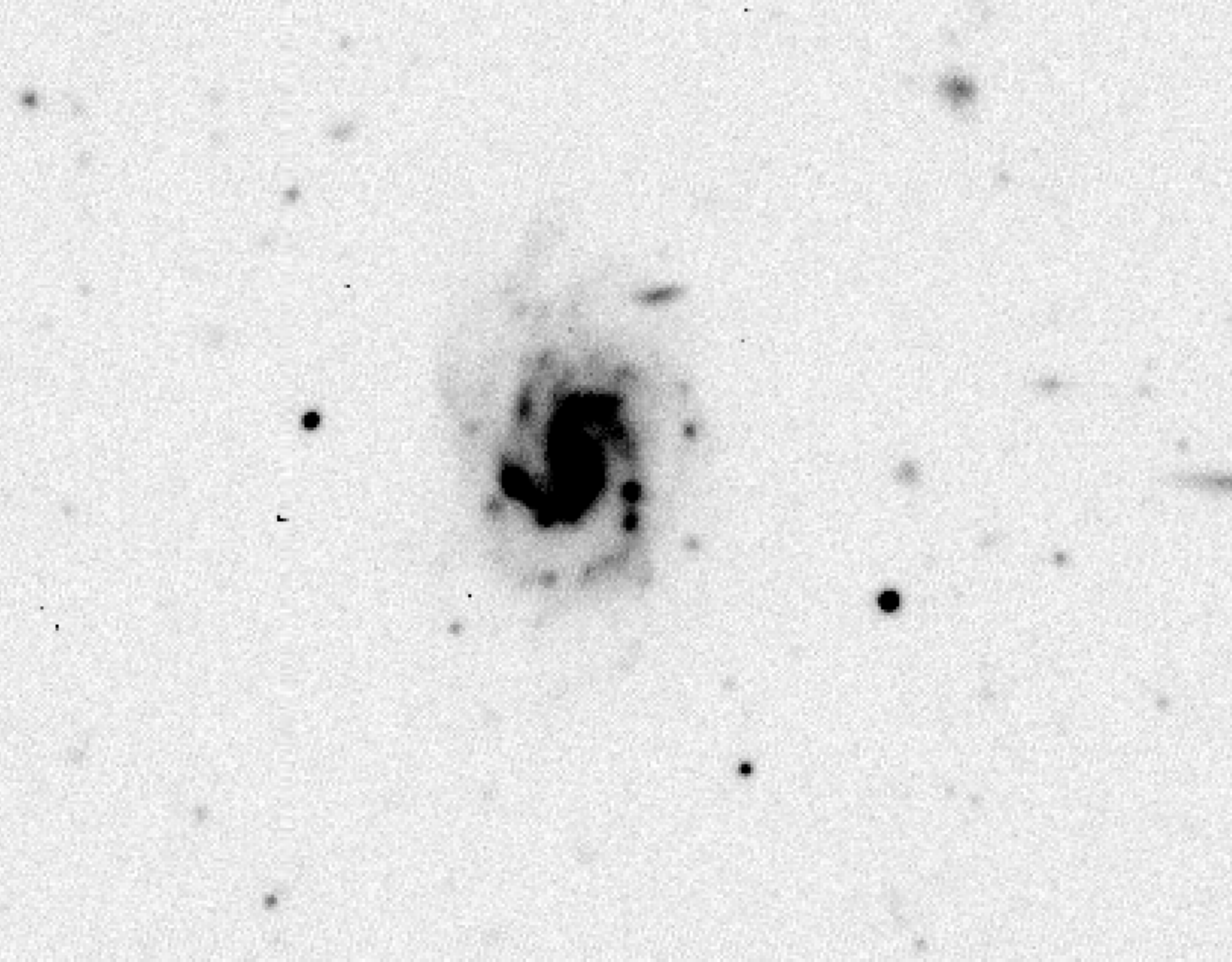}\hfill\hspace{-0.4cm}\includegraphics[scale=0.3]{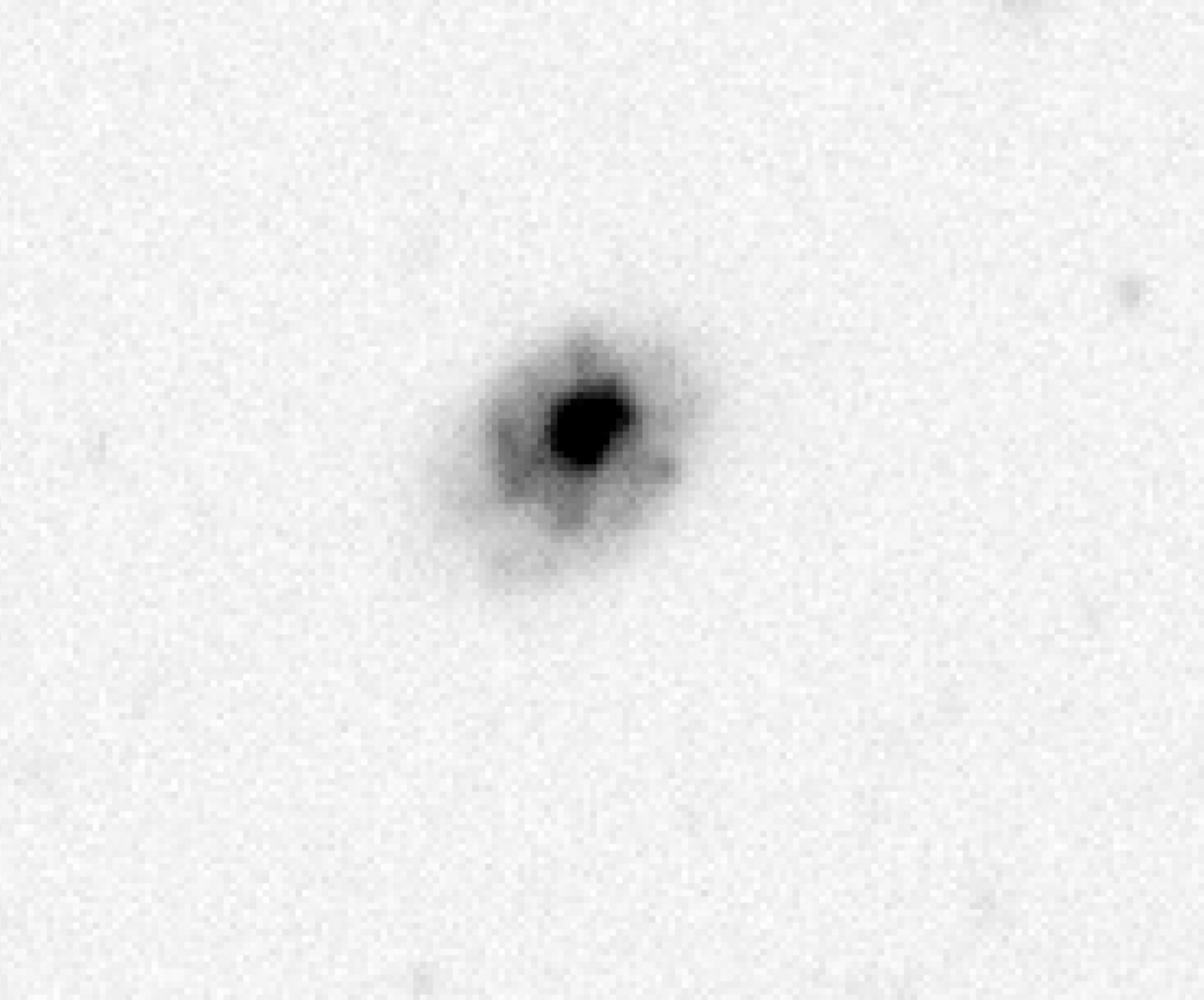}}
\vspace{-0.5cm}
\setcounter{figure}{2}
\caption{Two examples of each JClass=4, 3, 2 and 1 (from top to
  bottom) PM2GC stripping candidates. They belong to all types of
  environmental conditions found in the PM2GC, from groups to single systems. 
}
\end{figure*}

\begin{table}
\scriptsize
\caption{PM2GC stripping candidates.}
\begin{center}
\begin{tabular}{lllccllclccl}
 MGC ID & RA & DEC &  JClass &  z & Comments \\
&&&&& \\ 
 648     & 150.3656  & 0.15526  &1  &    0.06646  & \\
 669     & 150.50264 & 0.17896  &1  &    0.04648  & \\
 738     & 150.36314 & -0.04522 &1  &    0.06571  & \\
 954     & 150.51414 & -0.21385 &2  &    0.0452   & tidal \\
 1158   & 151.05859 & 0.26304  &1  &    0.0671   & \\
 1241   & 150.82014 & -0.00452 &1  &    0.06571  & \\
3425   & 153.00465 & -0.08858 &1  &    0.06978  & \\
3481   & 152.71695 & -0.13429 &1  &    0.06882  & \\
3924   & 153.24818 & -0.07149 &1  &    0.06285  & \\
3984   & 153.52425 & -0.12731 &4  &    0.04649  & \\
4686   & 153.89871 & -0.26982 &2  &    0.06341  &  tidal \\
4946   & 154.6284  & 0.08481  &2  &    0.06235  & \\
5055   & 154.53592 & -0.08383 &2  &    0.06111  & \\
5215   & 154.24265 & -0.248   &2  &    0.06296  & \\
5789   & 155.41989 & 0.21857  &1  &    0.05846  & \\
7591   & 156.81654 & -0.20765 &1  &    0.05633  & tidal \\
8694   & 158.45477 & 0.16144  &1  &    0.05557  & interaction \\
8721   & 158.53624 & 0.00101  &3  &    0.065    & \\
8770   & 158.53114 & -0.04816 &1  &    0.06486  & tidal \\
9223   & 158.97345 & -0.05658 &1  &    0.0587   & tidal \\
11695  & 161.56158 & 0.05042  &4  &    0.0466   & merger \\
12660  & 162.84444 & 0.2626   &1  &    0.04004  & prob tidal \\
13024  & 162.88519 & -0.28004 &1  &    0.05635  & \\
13384  & 163.26303 & -0.22522 &1  &    0.05133  & \\
13515  & 163.95135 & 0.22002  &2  &    0.04109  & tidal \\
14272  & 164.21922 & 0.02732  &1  &    0.05488  & \\
15672  & 166.44359 & 0.23992  &1  &    0.06717  & \\
17048  & 168.15927 & 0.13376  &3  &    0.0489   & tidal \\
17873  & 168.15373 & 0.0109   &1  &    0.05516  & tidal \\
17945  & 168.86038 & 0.2699   &1  &    0.04402  & \\
18060  & 168.74698 & -0.01211 &1  &    0.04302  & \\
19313  & 169.72412 & 0.02301  &1  &    0.05846  & \\
19482  & 170.63046 & -0.01728 &1  &    0.04078  & \\
20159  & 171.09225 & -0.27661 &2  &    0.04906  & \\
20769  & 171.82335 & 0.19009  &1  &    0.0491   & \\
20883  & 171.93929 & -0.1212  &2  &    0.06175  & \\
20925  & 172.13615 & -0.17166 &2  &    0.06421  & \\
24049  & 175.8248  & -0.18163 &1  &    0.05571  & \\
24069  & 176.01851 & -0.21116 &1  &    0.04844  & tidal \\
25500  & 177.90108 & 6.0E-4   &2  &    0.0605   & \\
26189  & 178.56143 & 0.01557  &1  &    0.05636  & \\
26597  & 179.07112 & 0.05586  &1  &    0.06486  & tidal \\
29909  & 182.80855 & 0.18497  &1  &    0.06299  & tidal \\
30102  & 182.64824 & -0.17155 &1  &    0.05052  & tidal \\
30802  & 183.97357 & 0.08151  &1  &    0.04021  & \\
31663  & 184.7652  & 0.0811   &1  &    0.06818  & tidal \\
36727  & 190.8569  & -0.08998 &1  &    0.04782  & \\
40457  & 195.38736 & -0.08069 &3  &    0.06799  & \\
42932  & 197.68634 & 0.03204  &1  &    0.04079  & \\
44092  & 199.55598 & -0.14834 &2  &    0.04877  & tidal \\
44601  & 199.73911 & -0.19856 &1  &    0.05426  & \\
45094  & 200.17049 & -0.20287 &1  &    0.0467   & tidal \\
45479  & 200.8947  & -0.13107 &1  &    0.05159  & \\
45979  & 201.23416 & -0.1341  &2  &    0.0666   & tidal \\
48127  & 204.03214 & 0.2167   &1  &    0.06204  & \\
48157  & 204.00653 & 0.26251  &1  &    0.06156  & \\
57255  & 212.15759 & -0.22513 &1  &    0.05209  & \\
57486  & 212.89355 & 0.16627  &1  &    0.05275  & \\
59348  & 214.51141 & 0.13088  &1  &    0.05451  & \\
59391  & 214.60939 & 0.16948  &1  &    0.05324  & \\
59597  & 214.42107 & -0.14422 &2  &    0.04964  & \\
60136  & 215.06152 & -0.07929 &1  &    0.05273  & \\
60151  & 214.69313 & -0.08358 &2  &    0.05329  & tidal/superposition?\\
61980  & 216.28696 & -0.12431 &1  &    0.05463  & tidal \\
62927  & 217.67888 & 0.25342  &3  &    0.05493  & prob merger \\
63504  & 218.13986 & 0.10742  &1  &    0.0554   & \\
63661  & 218.09081 & 0.17823  &2  &    0.05479  & \\
63692  & 218.00017 & 0.08422  &1  &    0.0557   & \\
63947  & 217.75774 & -0.18262 &2  &    0.05574  & \\
90126  & 163.0394  & 0.01269  &2  &    0.04057  & \\
95080  & 198.03625 & -0.23903 &1  &    0.04049  & \\
96244  & 214.64769 & 0.15756  &4  &    0.05303  & \\
96248  & 217.4342  & 0.08132  &1  &    0.05469  & \\
96328  & 214.13489 & 0.07133  &2  &    0.05259  & \\
96949  & 178.54276 & 0.13871  &3  &    0.05014  & interaction/projection \\
\end{tabular}
\tablecomments{Columns: (1) ID (2) MGC ID (3)  RA (J2000) (4) DEC (J2000) (5) JClass: from 5 (strongest)
  to 1 (weakest) (6) MGC redshift (7) comments. The table is published in its entirety in the 
 electronic edition of the journal.  A portion is 
 shown here for guidance regarding its form and content.}
\end{center}
\end{table}

\section{Location of stripping candidates}

Out of the 156+77 OMEGAWINGS+WINGS candidates with a spectroscopic
redshift, 107+55 ($\sim 70$\%) are cluster members.

Our clusters cover a wide range of $\sigma$ and $L_X$ (thus cluster
mass), but the number of candidates per cluster (spectroscopic
members plus possibly members because without a redshift) does not
depend on either of these observables, nor on redshift in our
narrow $z$ range (Fig.~4). This result does not change if we only
consider stripping candidates that are spectroscopically confirmed members.
Also considering only the number
of most secure candidates (JClass 3,4,5), there
is no sign of a correlation with $\sigma$ or $L_X$.
The independence of the number of candidates should be taken with
caution for two reasons: the incomplete spectroscopic membership, and
the fact that we are observing a fixed area on the sky (the OMEGACAM
field), that corresponds to different fractions of the cluster virial radius.

Nonetheless, it is striking that stripping candidates are found in all
clusters inspected and, even more, that they can be
present in large numbers even in 500 $\rm km \, s^{-1}$, low-mass
clusters. Moreover, the strength of the stripping signatures
is not correlated with $\sigma$ or $L_X$ either. In particular, the fraction of JClass=3,4,5 candidates
with respect to all
candidates is not significantly correlated with $\sigma$ or $L_X$.
This raises the question what is the minimum halo
mass that can trigger the stripping phenomenon, which we will address
in the following with the PM2GC sample.

While the average number of candidates per cluster that are members or could be
members (have no redshift) is 4.6$\pm 0.7$, this number rises to
6.1$\pm 0.5$ in
clusters of the Shapley supercluster, suggesting the supercluster
environment could be particularly favourable for stripping
phenomena. This is consistent with the hypothesis that ram pressure
effects are enhanced wherever the merging of structures produces
shocks and strong temperature gradients in the IGM (Owers et al. 2012,
Vijayaraghavan \& Ricker 2013).
There are smaller superclusters and cluster mergers in our sample, and
the relation with large scale structure and cluster substructure will be explored in forthcoming papers.

Candidates that are cluster members are observed at all clustercentric
radii, though their distribution is skewed towards larger radii than
the global population of members  (Fig.~5), as it is expected for a
population of mainly late-type galaxies. Here and throughout the paper
clustercentric radii are projected radii on the plane of the sky, in
units of $R_{200}$, measured from the Brightest Cluster Galaxy.
We note that the OmegaCAM field of view probes between $\sim$1.2 and 2.4 times the cluster 
virial radius $R_{200}$, depending on cluster redshift and extension,
thus in the left panel of Fig.~5 the coverage is complete for all
clusters only out to $r/r_{200}=1.2$. 

For about 80\% of the cluster member candidates, it is possible to identify one main
direction of the stripped material on the plane of the sky.
For these, the ``tentacles'' or main
tail point away from the cluster center in $\sim 35$\% of the cases,
point towards the cluster center in $\sim 13$\% of the cases, and form an
angle (not 0, nor 180 degrees) with respect to the cluster center
in $\sim 52\%$ of the cases. This non-alignment between the tails and
the direction to the cluster center can originate from non radial
orbits, but also if the stripping is caused by encounters with
IGM substructures and shocks.

Seven (4$\pm2$\%) of the OMEGAWINGS candidates that are not members
can be assigned to other structures (clusters or groups) along the
line of sight, in the foreground or background of the main WINGS
cluster in that field (flag=2 in Column 9 of Table~3). 
Some of the flag=2 candidates belong to groups with 
$\sigma < 400 \rm \, km \, s^{-1}$. The best examples are the 
two candidates in the field of A1069 (z=0.0651) that belong to a 
$\sigma =$372$\pm$84 $\rm \, km \, s^{-1}$ structure at $z \sim 0.56$
(Moretti et al. in prep.).

Finally, there is not sufficient spectroscopic information to
characterize the environment of the 42 remaining candidates with
redshifts (27$\pm4$\% of all candidates with redshift)
belonging neither to the main cluster nor to other fore-background
known structures (flag=0 in Table~3).

However, the group environment is better investigated with the PM2GC
sample.\footnote{The MGC 
stripe does not contain any X-ray cluster at z=0.04-0.07 
at the X-ray flux limit of the WINGS selection (BCS+eBCS +XBACS samples,
Ebeling et al. 1996, 1998, 2000), except A957x which has three
identified stripping candidates. Unfortunately, the MGC area covers only 
a fraction of the cluster, and the three candidates are outside of 
the PM2GC field.} The mass distribution of haloes hosting stripping
candidates is shown in Fig.~6. All PM2GC candidates are found in
haloes with masses $10^{11}-10^{14} M_{\odot}$.\footnote{Their hosting systems
can be ``groups'', binary or even single
systems according to the PM2GC classification (\S2.2). We stress that
such environmental definitions depend on the criteria chosen, which
are necessarily arbitrary at some level,  and therefore are only
roughly indicative of the richness of the host system. }
This is somewhat surprising, as the stripping phenomenon has always
been associated with ram pressure stripping in the past, and the
latter is often believed to be effective only in massive clusters with
a hot and dense intracluster medium. However, evidence for ram pressure
effects in groups is present in the literature (e.g. Rasmussen et
al. 2006, Sengupta \& Balasubramanyam 2006) and there is at least one
known case of ram pressure stripping in a galaxy pair, where NGC4485
is stripped during its passage through the extended HI distribution of
its companion, NGC4490 (Clemens et al. 2000).
The PM2GC sample has a considerable number of stripping candidates
with features as convincing as those in clusters
(JClass=4 and 3, see Fig.~3 and full sample online), and even they (as
all candidates) are clearly {\it not} located 
in a cluster. 

This is an interesting result, suggesting either that a) ram pressure
stripping can be efficient in lower mass haloes than commonly
believed (e.g. Clemens et al. 2000), or b) there are other types of
physical processes that work in groups (and perhaps clusters as well) that produce similar
debris morphologies and similar signatures for stripped gas. In
groups and low-mass haloes in general, the most likely candidates for such processes are 
tidal interactions and minor or major mergers. We reiterate that 40\%
of the PM2GC stripping candidates have been flagged as possibly tidal,
interacting, undergoing merging (no harassment candidates have been
found in the PM2GC).
Studying in detail a sample of stripping candidates in groups will therefore be crucial to
understand the impact of gas stripping processes on galaxy
evolution in general. An ongoing program of this kind based on the
sample presented in this paper is described in sec.~7.


\begin{figure}
\includegraphics[angle=0,scale=.70]{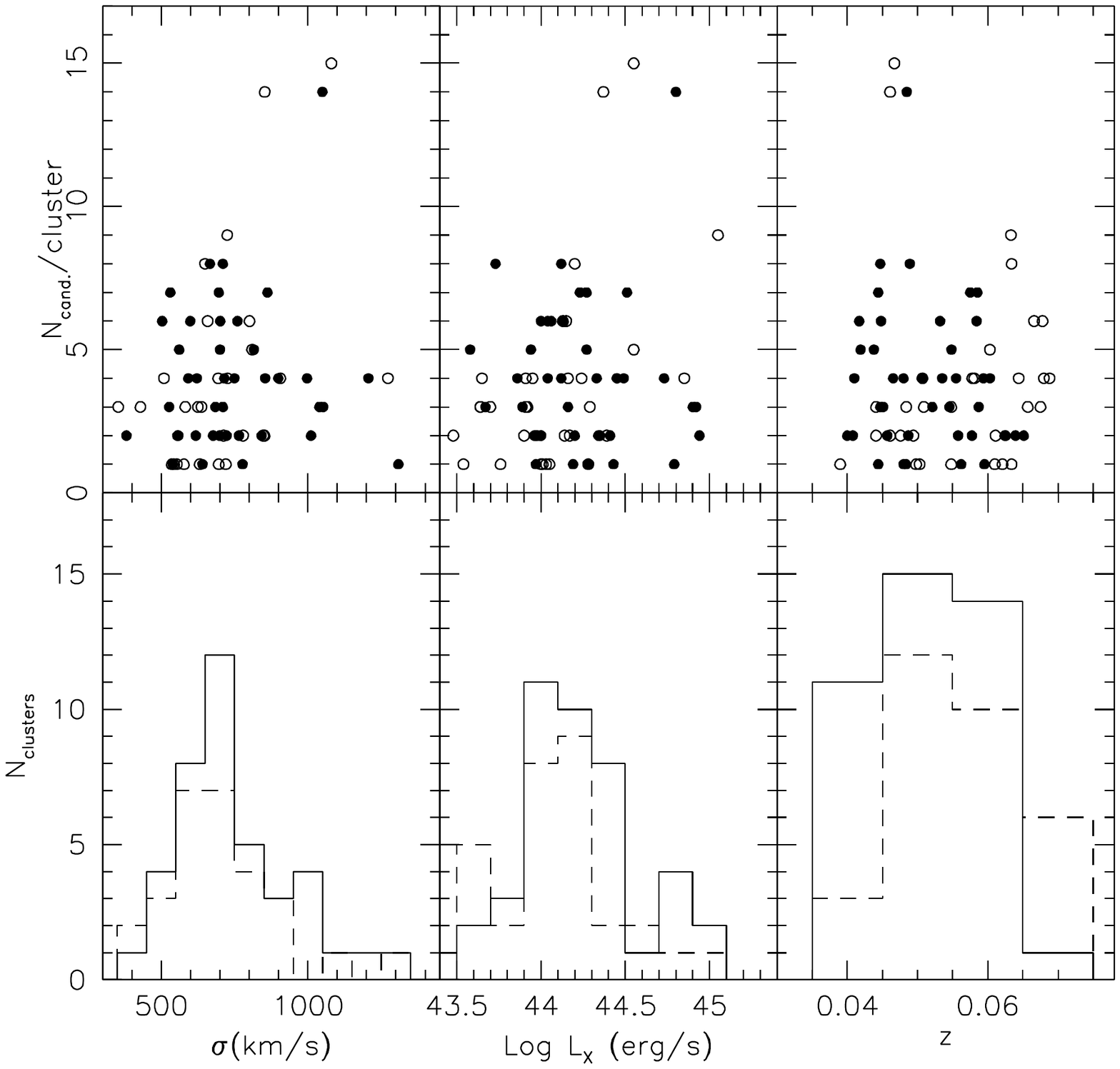}
\vspace{-3cm}
\caption{{\bf Top} Number of stripping candidates per cluster (members
  and possibly members, because without redshift) as a function of
  cluster $\sigma$, $L_X$ and
  $z$.
{\bf Bottom} Distribution of $\sigma$, $L_X$ and $z$ of clusters with
candidates.
Filled points and solid histogram OMEGAWINGS, empty points and dashed
histogram WINGS.
}
\end{figure}

\begin{figure}
\vspace{-3cm}
\includegraphics[angle=0,scale=.70]{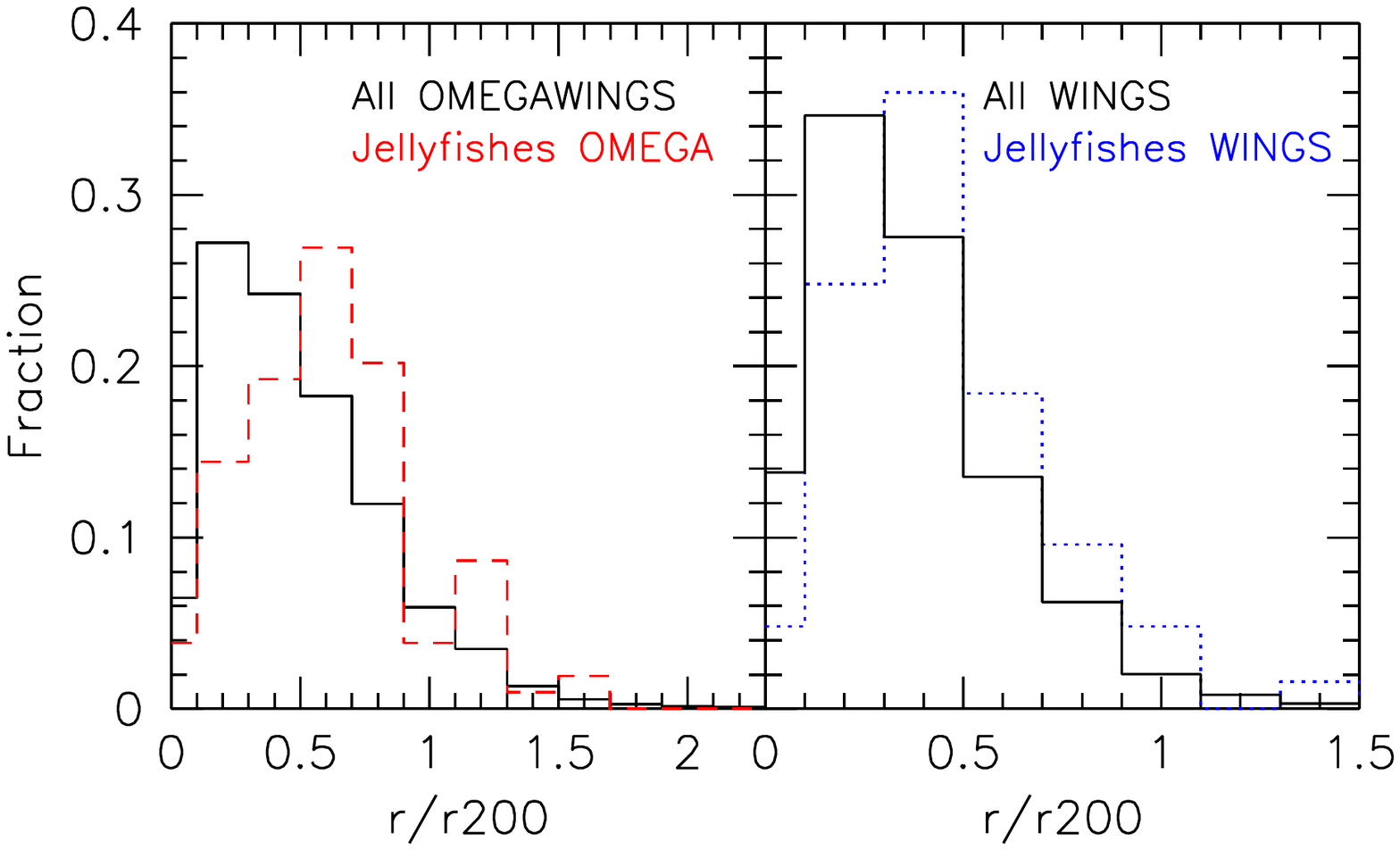}
\vspace{-3cm}
\caption{Radial distribution in units of $R_{200}$ of all OMEGAWINGS cluster members 
  (left, black histogram) plus OMEGAWINGS stripping candidates (left, dashed red histogram),
  and all WINGS cluster members (right, black histogram) plus WINGS 
  candidates (right, dotted blue histogram). 
}
\end{figure}

\begin{figure}
\vspace{-3cm}
\includegraphics[angle=0,scale=.70]{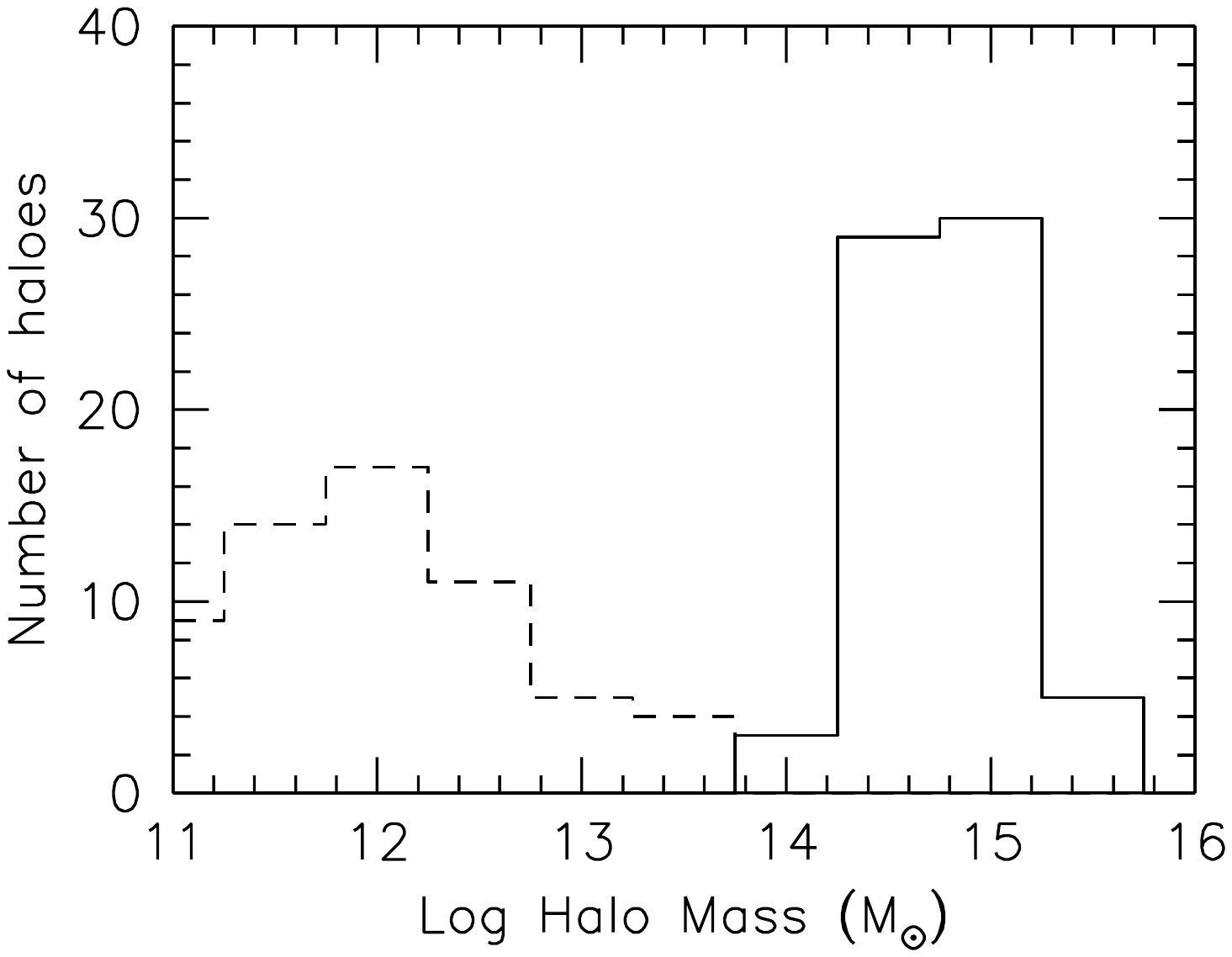}
\vspace{-2cm}
\caption{OMEGAWINGS+WINGS (solid histogram)
  and PM2GC (dashed histogram) mass distribution of haloes hosting
  stripping candidates. All JClasses are included.
}
\end{figure}

\section{Morphologies, star formation, colors and masses}

The distribution of morphological types for the three
samples\footnote{As written in sec.3.1, morphological classifications
  are available only for WINGS and PM2GC, while the analysis is still
  ongoing for OMEGAWINGS. The OMEGAWINGS morphologies in Fig.~7 are
  those taken from the WINGS images of OMEGAWINGS clusters.} of stripping candidates is shown in Fig.~7. 
The great majority of them are disk 
galaxies of types between Sab and Sc, with a tail of 
earlier and later types. According to a Kolmogorov-Smirnov (KS) test, 
while the OMEGAWINGS and WINGS distributions could be drawn from the
same parent population, the PM2GC distribution differs significantly
from the other two, with a KS probability greater than 99.8\% and 99.999\%
vs. OMEGAWINGS and WINGS, respectively.

We visually checked all candidates
classified by MORPHOT as ellipticals or S0s and, indeed, except for
the stripping signatures, they appear to have early-type morphologies.
When available, their spectra always show emission lines.

The bottom panel of Fig.~7 shows the morphological distributions of
the whole parent samples of galaxies in WINGS and PM2GC. They exhibit a
much more prominent early-type population and are
radically different from the morphological distributions of stripping candidates:
the KS test can rule out that each stripping candidates distribution
is drawn from its parent catalog with $>99.999$\% probability.

The SFR-stellar mass relation of candidates is contrasted
with that of all star-forming galaxies in Fig.~8 for OMEGAWINGS+WINGS
and PM2GC separately. Stripping candidates tend to
be located above the best fit to the relation, indicating a SFR excess
with respect to non-candidate galaxies of the same mass.
To make sure that this conclusion is not influenced by contamination
of tidally disturbed galaxies, we plot non-tidal and possibly tidal
candidates with different symbols.  Even considering only the most
secure candidates (non-tidal, of JClass 3,4,5), the SFR excess is
clearly visible in the plot.

If we define the fraction of secure stripping candidates as the ratio
between the number of candidates of JClass$\geq$3
and the total number of galaxies with a SFR$>0.1 M_{\odot}/yr$
that can be located on the SFR-Mass diagram above the mass
completeness limit of each sample,
this fraction is about 2\% for both OMEGAWINGS and PM2GC. This gives a
rough indication of the frequency of the most secure stripping candidates within the
global galaxy population.

The SFR excess, measured as distance from the fit at fixed mass, is
plotted in the bottom panels.
On average, the SFR of stripping candidates is enhanced by a factor 2.3/1.7 in
OMEGAWINGS candidates of JClasses (3,4,5)/(1,2), respectively  (red and
green points), and a factor $\sim 8.6$/1.7 in PM2GC, at masses above
the mass completeness limit.\footnote{The mass limits are computed as the mass
of the reddest galaxy at the highest redshift at the spectroscopic magnitude limit
in each sample, $\rm logM/M_{\odot}=9.8$ for WINGS and 10.2 for PM2GC (see Vulcani et al. 2011 and Calvi et al. 2013 for details).}
Comparing the least square linear fit of the SFR-mass relations of stripping candidates with
that of all other galaxies (keeping fixed the slope shown in Fig.~8),
the offset is 2.5 times the errorbar on the intercept, thus the excess
can be considered significant approximately at the 98.7\% level.

The individual SF estimates are highly uncertain as they are obtained extrapolating
the SF rate measured within the central galaxy regions covered by the
fibre to a total value assuming a constant mass-to-light ratio (see
\S3). To assess their reliability as integrated SFR estimates, we have
derived SFRs from W4 fluxes from the ALLWISE Source Catalog 
(Wright et al. 2010, Mainzer et al. 2011)
using eqn.(14) in Rieke et al. (2009)
and rejecting those sources that are flagged as spurious detections or
image artifacts. Comparing the WISE-based SFR-mass relation of candidates 
and non candidates (not shown), we derive the integrated SFR excess 
(dashed lines in the bottom left panel of Fig.~8). Due to the
relatively high SFR detection limit of WISE ($\sim 1 M_{\odot}
yr^{-1}$ at the WINGS redshifts) the statistics are poor, but qualitatively the WISE
estimates confirm our spectral modeling findings: on average, using WISE,
the SFR of stripping candidates is enhanced by a factor 1.8/1.7 for JClasses (3,4,5)/(1,2).


\begin{figure}
\vspace{-3cm}
\includegraphics[angle=0,scale=.85]{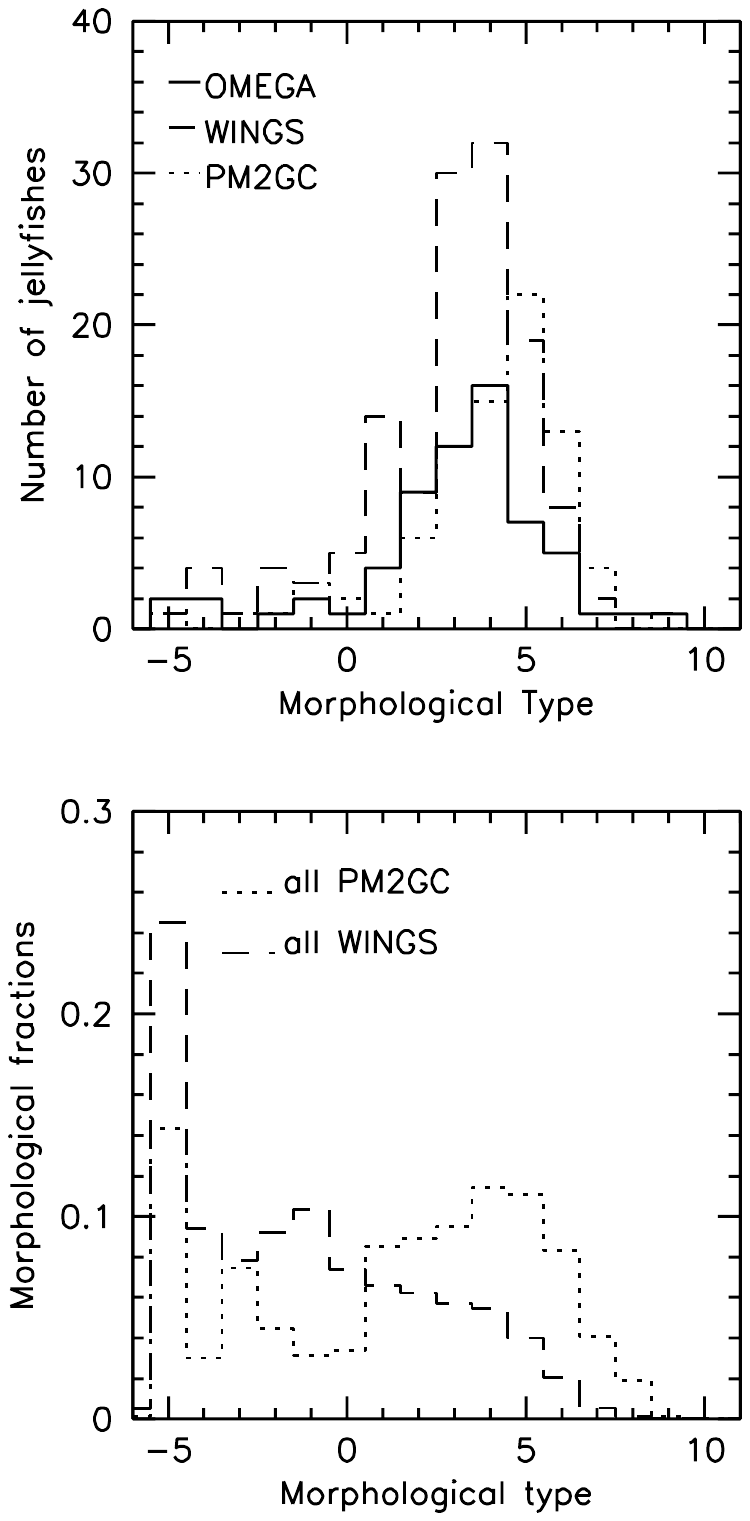}
\vspace{-3cm}
\caption{Top panel. Distribution of morphological types for stripping candidates
  in OMEGAWINGS (solid histogram), WINGS (dashed histogram) and PM2GC
  (dotted histogram). The main morphological types are: -5 =
  elliptical, -2 = S0, 1 = Sa, 2= Sab, 3 = Sb, 4=Sbc, 5 = Sc, 7 = Sd, 9 = Sm.
Bottom panel. As in top panel, distribution of morphological types
for all galaxies in the WINGS and PM2GC parent galaxy samples.
sample}
\end{figure}

\begin{figure*}
\vspace{-1cm}
\centerline{\includegraphics[scale=0.5]{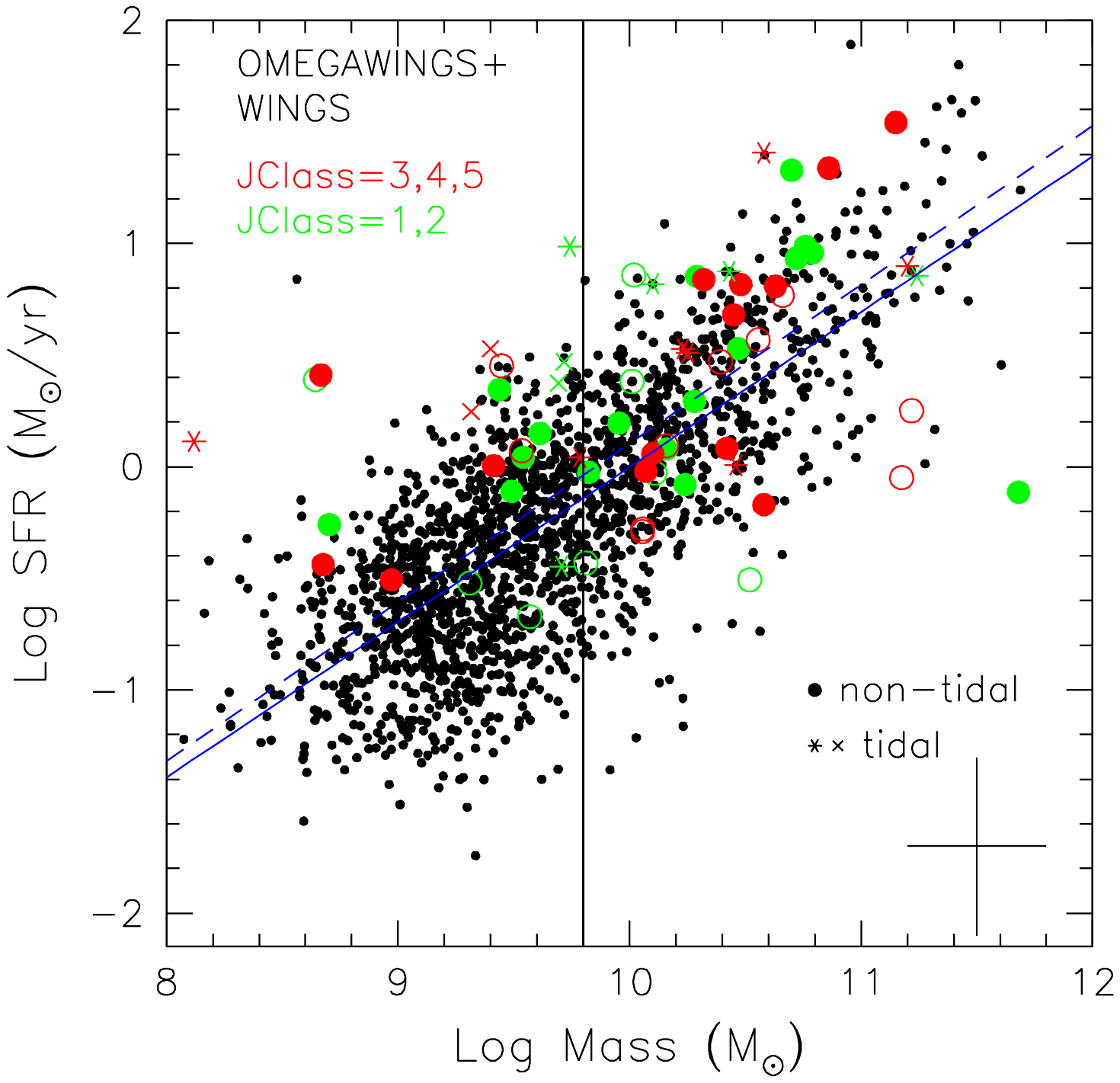}\hspace{-2cm}\includegraphics[scale=0.5]{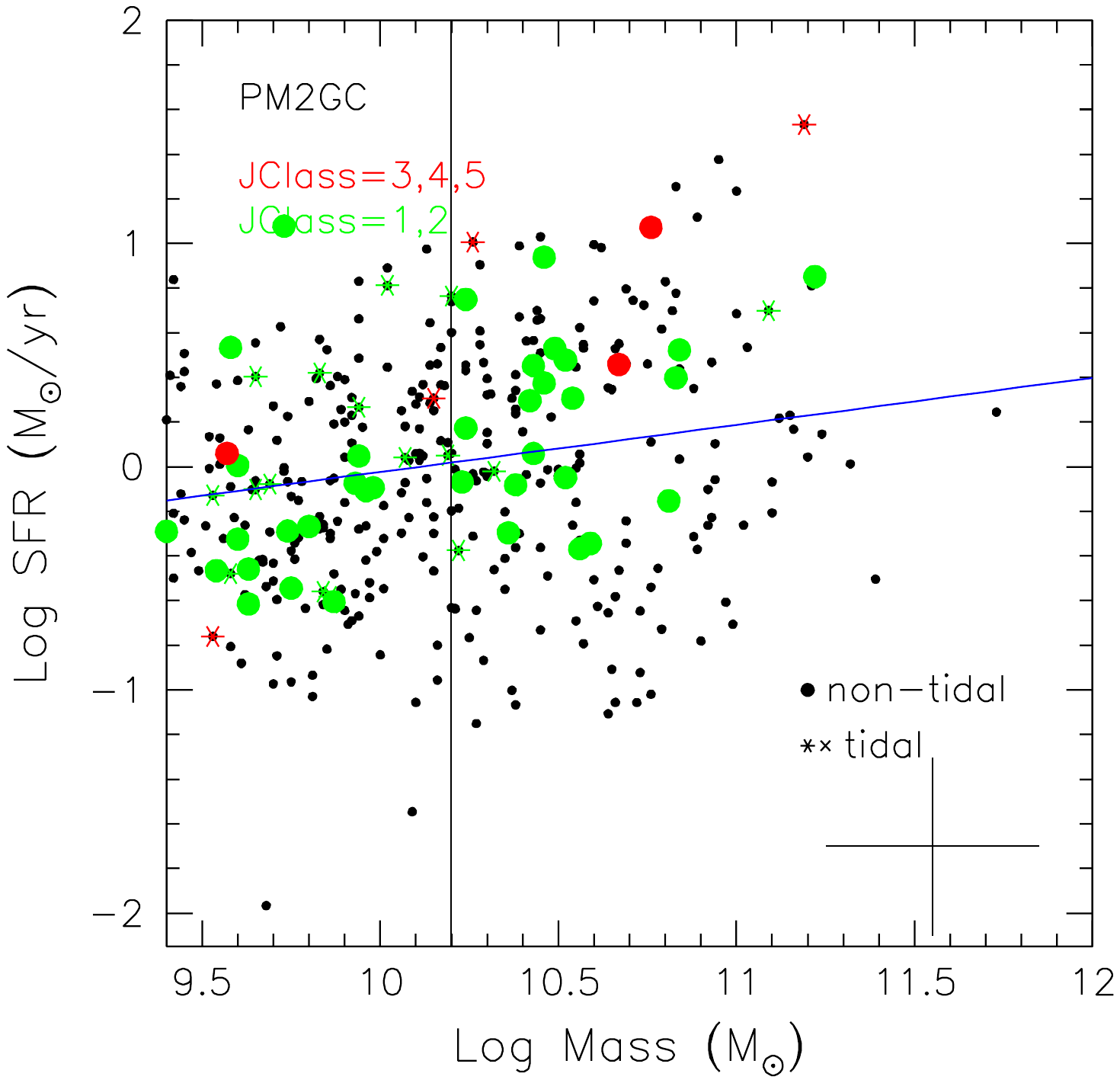}}
\vspace{-8cm}
\centerline{\includegraphics[scale=0.5]{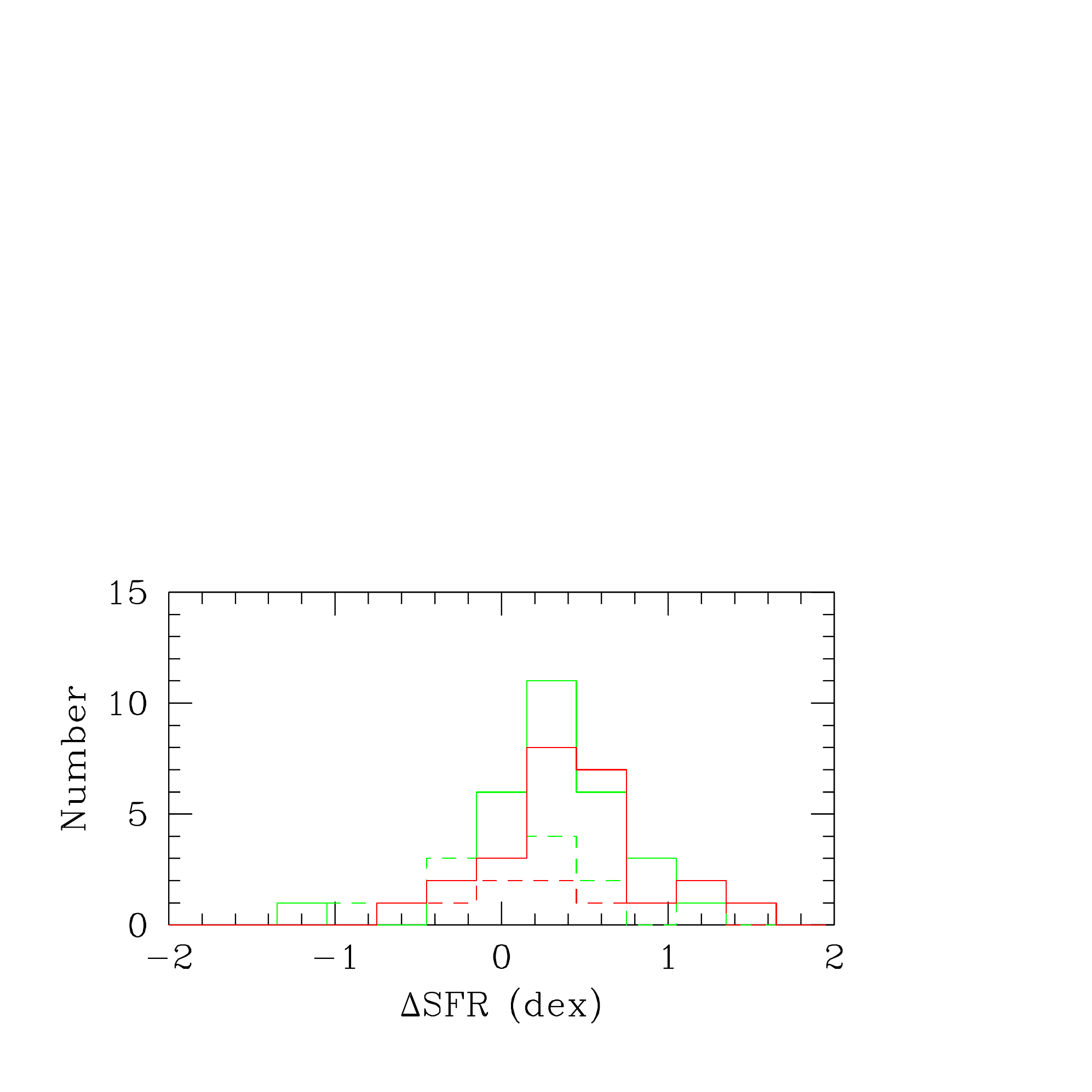}\hspace{-2cm}\includegraphics[scale=0.5]{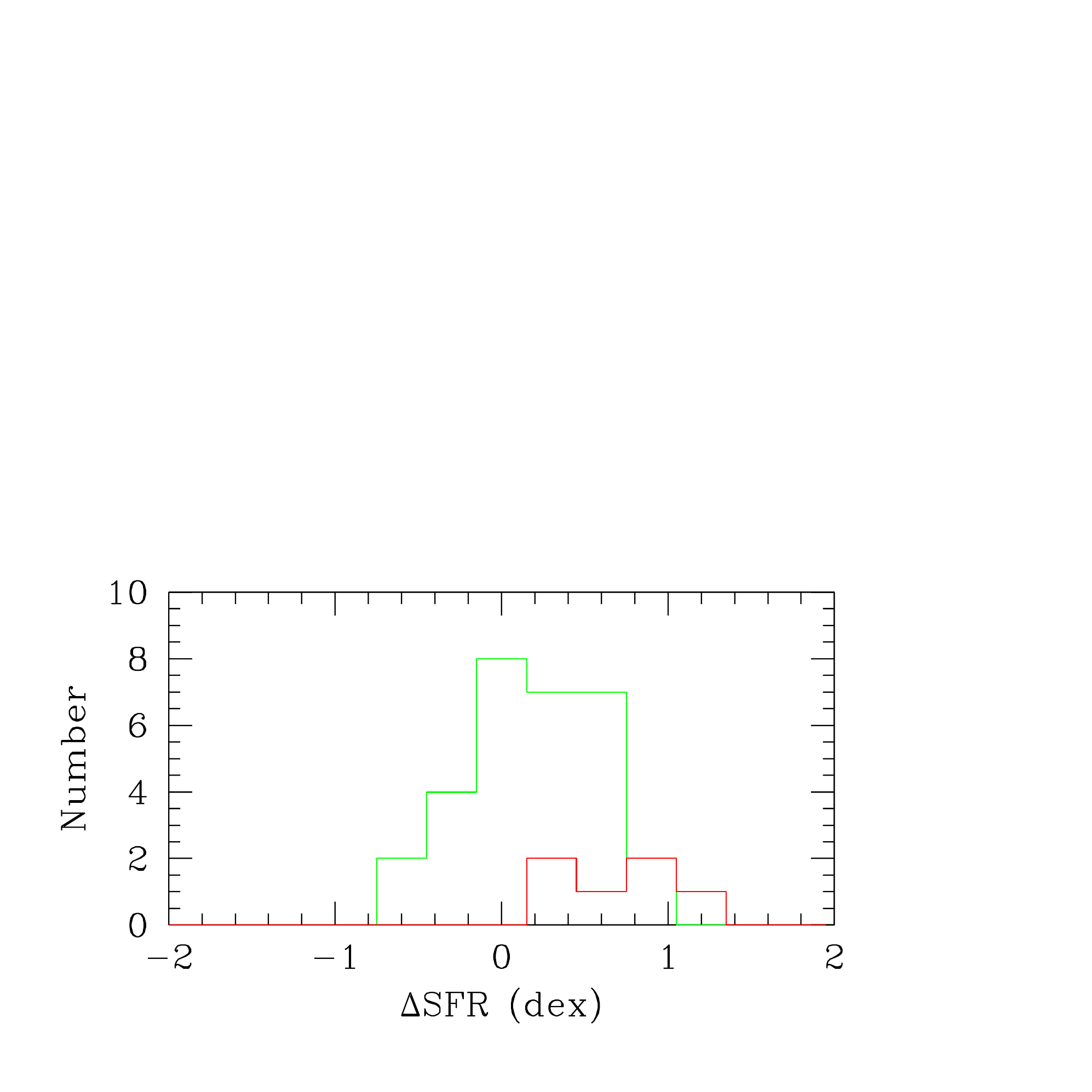}}
\vspace{-1cm}
\caption{{\bf Top} SFR-Mass relation for stripping candidates and all other galaxies in 
  OMEGAWINGS+WINGS (left) and PM2GC (right). JClass=1 and 2 in green,
  3,4 and 5 in red. Filled circles are non-tidal OMEGAWINGS and PM2GC, stars are
  possibly tidal OMEGAWINGS and PM2GC. Empty circles are non-tidal WINGS,
  crosses are possibly tidal WINGS. The blue line is the least square fit of cluster 
  (solid line) and WINGS field (dashed line) galaxies in the left panel, and of all PM2GC 
  galaxies in the right panel. The vertical lines indicate the mass
  completeness limit of each survey.
The uncertainty on the SFR can be estimated as 
the scatter obtained using independent SFR estimates of galaxies in our 
sample. Comparing with SDSS and WISE values, this 
uncertainty turns out to be $\sim 0.4$ dex and is shown in the right
bottom corner of the plot.
{\bf Bottom} Distribution of SFR excess with respect to the best fit 
SFR-Mass relation for OMEGAWINGS (left) and PM2GC (right). JClass=1 and 2 in green,
  3,4 and 5 in red. In the left panel, solid lines are for the model 
  SFR estimates, dashed histograms for WISE SFR estimates (see
  text). Errorbars as in the left panel.
}
\end{figure*}

\begin{figure}
\includegraphics[angle=0,scale=.70]{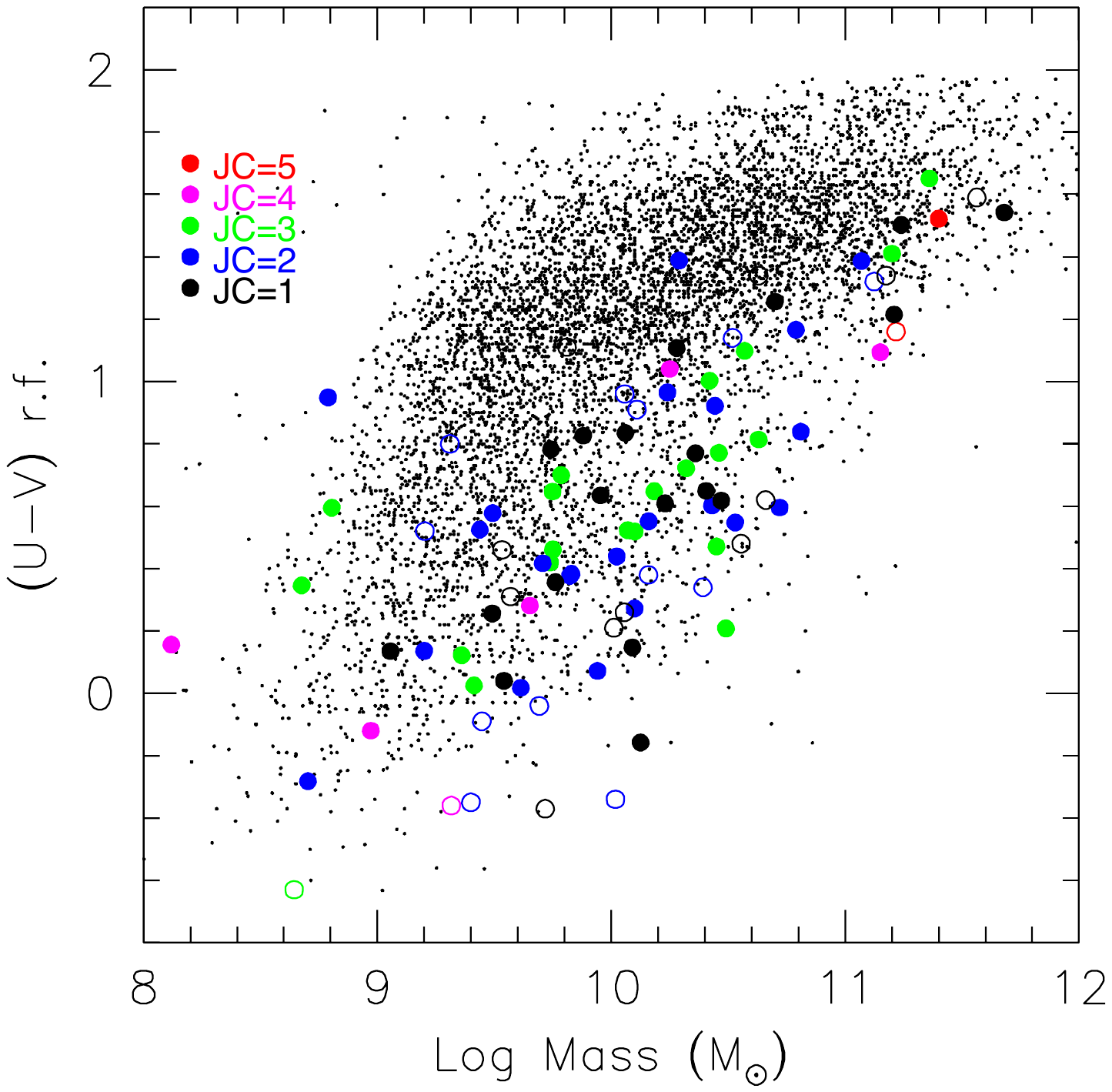}
\vspace{-3cm}
\caption{U-V rest frame color-mass diagram for all members of 
  OMEGAWINGS clusters (small black points) and for stripping candidates of different 
  JClasses= 5 red; =4 magenta; =3 green; =2 blue; =1 black. 
OMEGAWINGS=filled symbols, WINGS=empty symbols. 
}
\end{figure}

Considering the spectral classes, the great majority of stripping
candidates have emission lines. Only 3 out of 85 galaxies with an
assigned spectral type in OMEGAWINGS are k+a's, and only 2 are k's.
Similar trends are found in WINGS (no k+a, 2 k's out of 27) and PM2GC
(1 k+a and 6 k's out of 67). 
The lack a significant  ongoing SFR in k+a's and k's candidates, not
just in the galaxy center but throughout the galaxy,
is confirmed by the WISE data, that find no detection or very weak 
SFR upper limits ($<< 1 M_{\odot}/yr$). 
%
%

All k+a and k stripping candidates have weak stripping signatures (mostly JClass 1 or 2).
This indicates that, generally, the phase when the stripping is most
evident in the optical image corresponds to an early stage
of the process, when the galaxy SFR is enhanced, probably before being quenched
in later phases. 

The rest frame U-V-stellar mass relation is shown in
Fig.~9 for OMEGAWINGS+WINGS.\footnote{The color is not available for
PM2GC.}  Stripping candidates are among the bluest galaxies of their
mass, and are mostly located in the blue cloud, but they could not be singled out simply on
their location in the color-mass diagram. Their color does not
depend on JClass.

Finally, our stripping candidates cover a wide range in stellar mass, from $\rm
log M/M_{\odot} < 9 $ to $> 11.5$, and there is no correlation
between mass and JClass. Their stellar mass distribution,
both in clusters and in the field, is similar to that of the
global galaxy population in their environment (Fig.~10): for all three
samples, a KS test is unable to reject the hypothesis that they are drawn from the same
parent population.

\begin{figure}
\includegraphics[angle=0,scale=.70]{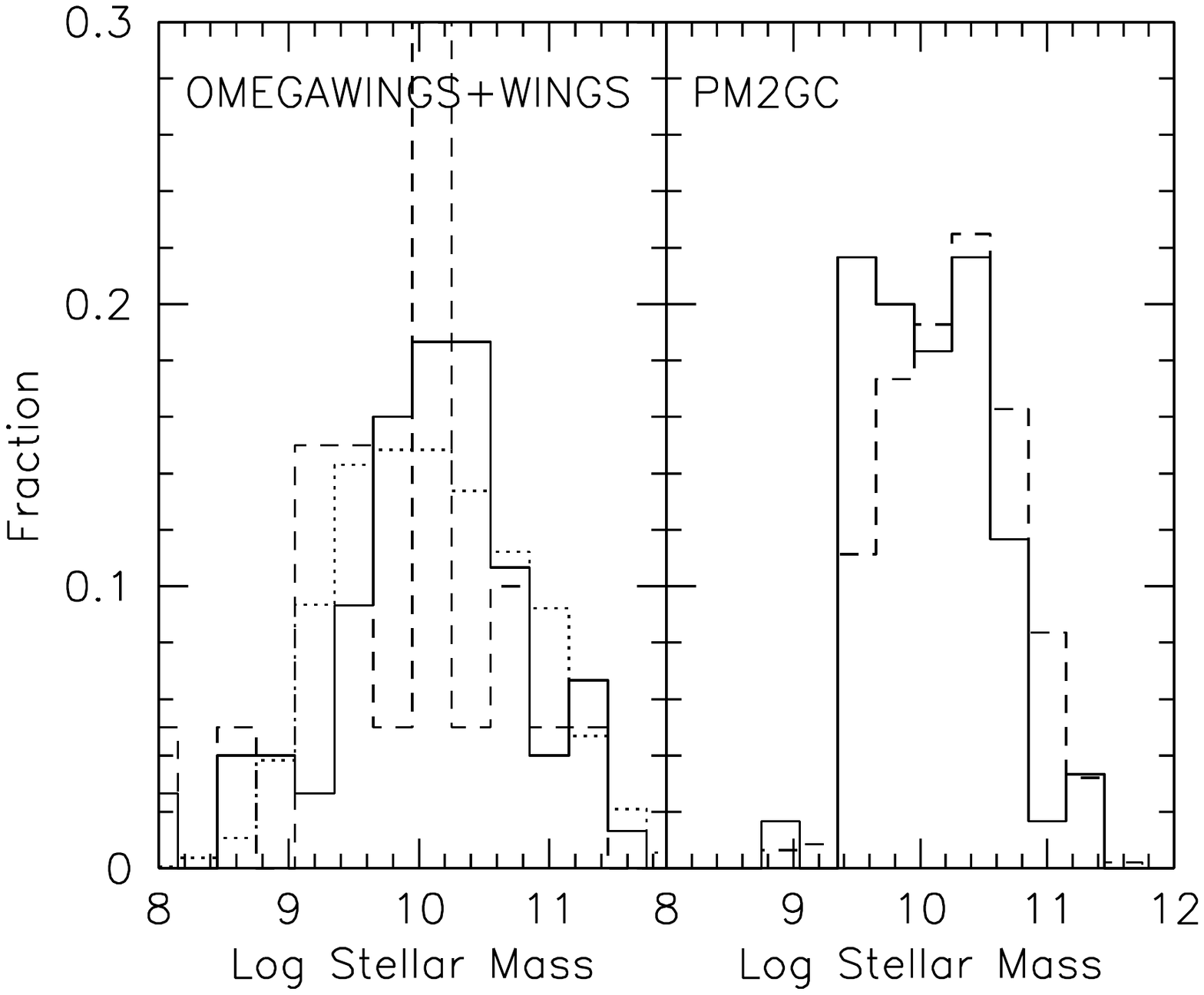}
\vspace{-3cm}
\caption{Galaxy stellar mass fractional distribution. {\bf (Left) } OMEGAWINGS
  candidates (solid line), WINGS candidates (dashed line) and all
  cluster members in the OMEGAWINGS+WINGS database (dotted line).
  {\bf (Right)} PM2GC candidates (solid line) and all galaxies (dashed line).
}
\end{figure}

\section{Conclusions}
Jellyfish galaxies are galaxies that
exhibit tentacles of material that appear to be stripped from the
galaxy and are the most extreme examples of galaxies in the process
of being stripped of their gas.
We have searched for galaxies whose morphology is suggestive
of gas-only removal mechanisms, from extreme jellyfishes to
less spectacular examples with evidence of debris tails and
morphologies of stripped material or asymmetry suggestive of unilateral external
forces. This is the first systematic search for such signatures at low
redshift. The purpose of this atlas is to provide a large sample
suitable for statistical and follow-up studies
that will be able to securely identify the process responsible for the
observed morphologies.

This paper presents the largest sample of stripping candidates known
to date: 344 galaxies in 71 galaxy clusters of
the OMEGAWINGS+WINGS sample, and 75 candidates in groups and
lower mass structures in the PM2GC sample, all at $z=0.04-0.07$.
Stripping candidates have been visually selected on the basis of B or
V deep images.
We present the atlas of single filter and (when available) color-composite images
of all candidates, together with catalogs of positions, redshifts,
JClass and eventual cluster membership.

Stripping candidates have been found in all clusters inspected,
that have $\sigma$ ranging from $\sim 500$ to $\sim 1200 \, \rm km \, s^{-1}$.
The number of candidates per cluster does not depend on the cluster $\sigma$
or $L_X$, and candidates are found at all clustercentric radii, out
to $\sim 1.5$ times the cluster virial radius that is the area covered
by our imaging in these clusters.
An in-depth analysis of the environments of these galaxies 
will be presented in a separate paper
 (Jaff\'e et al. in prep.).

While all jellyfishes previously known from the literature are in
clusters (but see e.g. Clemens et al. 2000 for a well studied case of
ram pressure stripping in a pair), we find striking stripping
candidates (highest JClasses) also outside of clusters, in groups and lower mass haloes
of the PM2GC sample, with masses in the range $10^{11}-10^{14}
M_{\odot}$. However, especially for the lowest JClasses, it is
possible that the observed features originate from
other mechanisms, such as minor or major mergers and
tidal interactions.
This result deserves further investigation, to fully understand the
role and the cause of gas stripping in groups and its impact on
galaxy evolution in general and on the global quenching of star formation.

Our preliminary analysis shows that the star formation rate is enhanced on average by a factor
of 2 in stripping candidates compared to non-candidates of the same
mass (2.5$\sigma$). This suggests that the process responsible for the stripping causes a
significant increase in the star formation activity.
There are a few non-starforming candidates (5-10\%), either in a
post-starburst phase or spectroscopically passive, and they display
rather weak stripping signatures. 
Our sample comprises candidates of all masses, from 
$\rm log \, M/M_{\odot} < 9 $ to $\rm > 11.5 M_{\odot}$, indicating that 
whatever causes the stripping phenomenon can be effective on galaxies of
any mass.

Only Integral Field spectroscopic observations can fully reveal
the cause and effects of gas removal in these galaxies,
and allow us to measure the stripping timescale, quantify the 
amount of stars formed in the stripped gas, unanbiguously identify the physical 
process responsible for the gas outflow and directly study the effects 
on the evolution of the galaxy. 
Integral Field Spectroscopy with MUSE/VLT was obtained for two
of our OMEGACAM jellyfishes. These data spectacularly reveals the
emission lines ($\rm H\beta$, [OIII], NII, $\rm H\alpha$ and SII) associated with the
ionized gas in the trails, out to several tens of kpc from the galaxy. This gas is ionized from the massive
stars in the star-formation knots that are visible in the optical images.
These results will be presented in a forthcoming paper (Jaff\'e et al. in prep.).
A larger IFS study on a statistically significant subsample of our
candidates is about to start with MUSE (ESO Large Program 196.B-0578) and will unveil the rich physics and implications of the
stripping phenomenon in galaxies as a function of galaxy environment and galaxy mass.

\acknowledgments
We thank the referee, Prof. Harald Ebeling, for his constructive comments
and suggestions that helped us improve the clarity and contents of
the paper.
We thank Joe Liske, Simon Driver and the whole MGC team for making
their great dataset easily accessible, and Rosa Calvi for all her work
on the PM2GC.
This work is based on two GTO programs on VST. We thank Massimo
Capaccioli, Enrico Cappellaro, Pietro Schipani, Andrea Baruffolo and
the whole VST and OmegaCAM teams for making this research possible.
This work is based on data obtained with AAOmega on the AAT, and
we acknowledge the generous support to the OMEGAWINGS project from the
Australian Time Assignment Committee.
We acknowledge financial support from a PRIN-INAF 2014 grant.
YLJ acknowledges support by FONDECYT grant N.~3130476.
BV was supported by the World Premier International Research Center
Initiative (WPI), MEXT, Japan and by the Kakenhi Grant-in-Aid for
Young Scientists (B)(26870140) from the Japan Society for the Pro-
motion of Science (JSPS). 
This research has made use of the NASA/IPAC Infrared Science Archive
and the NASA/IPAC Extragalactic Database (NED),
which are operated by the Jet Propulsion Laboratory, California
Institute of Technology, under contract with the National Aeronautics
and Space Administration.
This publication makes use of data products from the Wide-field
Infrared Survey Explorer, which is a joint project of the University
of California, Los Angeles, and the Jet Propulsion
Laboratory/California Institute of Technology, and NEOWISE, which is a
project of the Jet Propulsion Laboratory/California Institute of
Technology. WISE and NEOWISE are funded by the National Aeronautics
and Space Administration.
Funding for SDSS-III has been provided by the Alfred P. Sloan
Foundation, the Participating Institutions, the National Science
Foundation, and the U.S. Department of Energy Office of Science. The
SDSS-III web site is http://www.sdss3.org/.
SDSS-III is managed by the Astrophysical Research Consortium for the
Participating Institutions of the SDSS-III Collaboration.

\clearpage




\begin{thebibliography}{}
\bibitem[]{} Ahn, C.P. et al. 2014, ApJS, 211, 17
\bibitem[]{} Balogh, M.L., Navarro, J., Morris, S., 2000, MNRAS, ApJ, 540, 113
\bibitem[]{} Barnes, J.E. \& Hernquist, L., 1992, ARAA, 30, 705
\bibitem[]{} Boselli, A. \& Gavazzi, G., 2006, PASP 118, 517
\bibitem[]{} Bournaud, F., Combes, F., Jog, C.J., 2004, A\&A, 418, L27
\bibitem[]{} Bravo-Alfaro, H., Cayatte, V., van Gorkom, J., Balkowski,
  C., 2001, A\&A, 379, 347
\bibitem[]{} Byrd, G. \& Valtonen, M.,  1990, ApJ, 350, 89
\bibitem[]{} Calvi, R., Poggianti, B.M., Vulcani, B., 2011, MNRAS, 416, 727
\bibitem[]{} Calvi, R. et al., 2013, MNRAS, 432, 3141
\bibitem[]{} Cava, A. et al. 2009
\bibitem[]{} Cayatte, V., van Gorkom, J., Balkowski, C., Kotanyi, C., 1990, AJ, 100, 604
\bibitem[]{} Chung, A., van Gorkom, J., Kenney, J.D.P., Crowl, H., Vollmer, B.,  2009, AJ, 138, 1741
\bibitem[]{} Clemens, M.S., Alexander, P., Green, D.A., 2000, MNRAS, 312, 236
\bibitem[]{} Cortese, L. et al.  2007, MNRAS, 376, 157
\bibitem[]{} Cox, T.J., Jonsson, P., Somerville, R.S., Primack, J.R.,
  Dekel, A., 2008, MNRAS, 384, 386
\bibitem[]{} Cowie, L.L. \& Songaila, A.,  1977, Nature, 266, 501
\bibitem[]{} Davies, R.D. \& Lewis, B.M., 1973, MNRAS, 165, 231
\bibitem[]{} Dekel, A, \& Birnboim, Y., 2006, MNRAS, 368, 2
\bibitem[]{} De Lucia, G. 2010, arXiv 1012.3326
\bibitem[]{} Driver, S. et al. 2005
\bibitem[]{} Ebeling, H. et al., 1996, MNRAS, 281, 799
\bibitem[]{} Ebeling, H. et al., 1998, MNRAS, 301, 881
\bibitem[]{} Ebeling, H. et al., 2000, MNRAS, 318, 333
\bibitem[]{} Ebeling, H. et al. 2014, ApJ, 781, 40 
\bibitem[]{} Fasano, G. et al. 2006, A\&A, 445, 805
\bibitem[]{} Fogarty, L.M.R. et al., 2012, ApJ, 761, 169
\bibitem[]{} Fritz, J., et al., 2007, A\&A, 470, 137
\bibitem[]{} Fritz, J., et al., 2011, A\&A, 526, 45
\bibitem[]{} Fritz, J., et al., 2014, A\&A, 566, 32
\bibitem[]{} Fumagalli, M. et al.,  2014, MNRAS, 445, 4335
\bibitem[]{} Giovanelli, R. \& Haynes, M.P.,  1985, ApJ, 292, 404
\bibitem[]{} Gullieuszik, M. et al., 2015, A\&A in press (arXiv 1503.02628) 
\bibitem[]{} Gunn, J.E. \& Gott, J.R.,  1972, ApJ, 176, 1
\bibitem[]{} Haynes, M., Giovanelli, R., Chincarini, G., 1984, ARA\&A, 22, 445
\bibitem[]{} Hester, J.A., et al., 2010, ApJ, 716, L14;
\bibitem[]{} Ho, I.-T. et al., 2014, MNRAS, 3894, 3910
\bibitem[]{} Hopkins, P.F., Cox, T.J., Younger, J.D., Hernquist, L.,  2009, ApJ, 691, 1168
\bibitem[]{} Jaff\'e, Y. et al., 2015, MNRAS, 448, 1715 
\bibitem[]{} Kenney, J.D.P., Koopmann, R.A., 1999, AnJ, 117, 181
\bibitem[]{} Kenney, J.D.P., van Gorkom, J., Vollmer, B., 2004, AJ, 127, 3361
\bibitem[]{} Kenney, J.D.P. et al., 2014, ApJ, 780, 119 
\bibitem[]{} Kenney, J.D.P., Abramson, A., Bravo-Alfaro, H., 2015, AnJ
  in press, arXiv 1506.04041 
\bibitem[]{} Larson, R.B., Tinsley, B.M., Caldwell, C.N., 1980, ApJ, 237, 692 
\bibitem[]{} Liske, J. et al., 2003, MNRAS, 344, 307
\bibitem[]{} Lotz, J.M., Jonsson, P., Cox, T.J., Primack, J.R., 2010a,  MNRAS, 404, 575 
\bibitem[]{} Lotz, J.M., Jonsson, P., Cox, T.J., Primack, J.R., 2010b,  MNRAS, 404, 590 
\bibitem[]{} Machacek, M., Jones, C., Forman, W.R., Nulsen, P., 2006, ApJ, 644, 155
\bibitem[]{} Mainzer, A.  et al. 2011, ApJ,  731, 53 
\bibitem[]{} Merluzzi, P. et al., 2013, MNRAS, 429, 1747
\bibitem[]{} Mihos, C., Hernquist, L., 1994, ApJ, 425, L13
\bibitem[]{} Moore, B. et al., 1996, Nature, 376, 613
\bibitem[]{} Moretti, A. et al., 2014, A\&A, 564, 138
\bibitem[]{} Nulsen, P.E.J., 1982, MNRAS, 198, 1007
\bibitem[]{} Omizzolo, A. et al., 2014, A\&A, 561, 111
\bibitem[]{} Owers, M. et al.,  2012, ApJ, 750, L23; 
\bibitem[]{} Poggianti, B.M. et al., 2006, ApJ, 642, 188
\bibitem[]{} Rasmussen, J., Ponman, T.J., Mulchaey, J.S., 2006, MNRAS, 370, 453
\bibitem[]{} Rasmussen, J. et al., 2008, MNRAS, 388, 1245
\bibitem[]{} Rawle, T.D. et al., 2014, MNRAS, 442, 196
\bibitem[]{} Rieke, G.H. et al., 2009, ApJ, 692, 556
\bibitem[]{} Sengupta, C. \& Balasubramanyam, R., 2006, MNRAS, 369, 360
\bibitem[]{} Smith, R.J. et al., 2010, MNRAS, 408, 1417
\bibitem[]{} Sun, M. et al., 2005, ApJ, 619, 169
\bibitem[]{} Sun, M. et al., 2006, ApJ, 637, L81
\bibitem[]{} Valentinuzzi, T. et al., 2009,A\&A, 501, 851
\bibitem[]{} Varela, J. et al., 2009, A\&A, 497, 667
\bibitem[]{} Verdes-Montenegro, L. et al., 2001 A\&A, 377, 812
\bibitem[]{} Veilleux S., Cecil, G., Bland-Hawthorn, J., 2005, ARAA,
  43, 769 
\bibitem[]{} Vijayaraghavan, R., Ricker, P.M., 2013, MNRAS, 435, 2713
\bibitem[]{} Vollmer, B., 2003, A\&A, 398, 525
\bibitem[]{} Vulcani B. et al., 2011, MNRAS, 412, 246
\bibitem[]{} Yagi, M. et al., 2010, AJ, 140, 1814 
\bibitem[]{} Yoshida, M., et al., 2008, ApJ, 688, 918
\bibitem[]{} Williams, B.A. \& Rood, H.J., 1987, ApJS, 63, 265
\bibitem[]{} Wright, E.L. et al., 2010, AJ, 140, 1868 
\end{thebibliography}
\end{document}